\documentclass{cta-author}

{}
{}
{}
\usepackage{ragged2e}
\usepackage{graphicx}
\usepackage{float}
\usepackage{subcaption}
\usepackage{amsthm}
\usepackage{color}
\usepackage{tabulary}
\usepackage{verbatim}
\usepackage{enumerate}
\usepackage{rotating}
\usepackage[mathscr]{eucal}
\usepackage{ifthen}
\usepackage{makeidx}
\usepackage{multirow}
\usepackage{amsmath} 
\usepackage{amssymb}
\usepackage{booktabs}
\usepackage{textcmds}
\usepackage{nomencl}
\usepackage{verbatim}
\usepackage{rotating} 
\usepackage[mathscr]{eucal}
\usepackage{ifthen}
\usepackage{makeidx}
\usepackage{multirow}
\usepackage{amsmath} 
\usepackage{enumerate}
\usepackage{amsfonts}
\usepackage{amsmath}
\usepackage{hyperref}
\usepackage{algorithmic}
\usepackage{algorithm}

\begin{document}

\supertitle{}

\title{Measurement-Based Wide-Area Damping of Inter-Area Oscillations based on MIMO Identification}

\author{\au{Abilash Thakallapelli$^{1}$}, \au{Sukumar Kamalasadan$^{1\corr}$}}

\address{\add{1}{Electrical and Computer Engineering, UNC Charlotte, Charlotte, USA}
\email{skamalas@uncc.edu}}

\begin{abstract}

Interconnected power grid exhibits oscillatory response after a disturbance in the system. One such type of oscillations, the inter-area oscillations has the oscillation frequency in the range of 0.1 to 1 Hz. The damping of inter-area oscillations is difficult with local controllers, but it can be achieved using a Wide Area Damping Controller (WADC). For effective control, the input to the WADC should be the most observable signal and the WADC output should be sent to the most controllable generator. This paper presents a measurement-based novel algorithm for multi-input-multi-output (MIMO) transfer function identification of the power system based on optimization to estimate such oscillation frequencies. Based on the MIMO transfer function the optimal control loop for WADC is estimated. The WADC design is based on the discrete linear quadratic regulator (DLQR) and Kalman filtering for damping of inter-area oscillations. Since the MIMO identification is based on actual measurements, the proposed method can accurately monitor changes in the power grid whereas the conventional methods are based on small-signal analysis of a linearized model which does not consider changing operating conditions. The overall algorithm is implemented and validated on a RTDS/RSCAD\textsuperscript{\tiny\textregistered} and MATLAB\textsuperscript{\tiny\textregistered} real-time co-simulation platform using two-area and IEEE 39 bus power system models. 

\end{abstract}

\maketitle
\vspace{-10mm}
\section{Introduction}\label{sec1}
The increase in penetration of renewable energy sources and deregulation made the power system to operate near its rated operating limits. Under these circumstances, if there is a disturbance in the system this could affect the power system reliability and security. The oscillatory response which arise due to such disturbances have to be damped effectively and these oscillations are of two types 1) Local oscillations, and 2) Inter-area oscillations. The local oscillations have frequencies greater than 1 Hz and inter-area oscillations frequency lies in the range of 0.1 to 1 Hz \cite{ref101}. During inter-area oscillations generators in one coherent area swing together against other area generators \cite{chow2014power}. In general, such oscillations are suppressed using supplementary control techniques likes power system stabilizers (PSS) \cite{ref103}, flexible alternating current transmission systems (FACTS) devices \cite{ref105} and high-voltage direct current (HVDC) links \cite{ref104}. However, such controllers are effective only during local oscillations. As the inter-area oscillations may destabilize the power system \cite{ref102} it should be damped effectively.

In literature, several methodologies are presented to damp inter-area oscillations using a wide area control loop which is effective to damp an inter-area mode of interest. The wide area control loop is based on the most observable control signal which is input to the wide-area controller and the output of the controller is sent to the most controllable generator. This input/output combination can be estimated based on residue analysis \cite{refresidue1,refresidue2,refresidue3}, combination of residue and relative gain array (RGA) method \cite{refcom1}, and a joint observability-controllability measure \cite{refgeo1,refa4}. In all these  methods, the control-loop selection is based on linearizing the system at an operating point and analyzing the state space matrices obtained thereafter. If the system operating point changes then the control-loop obtained at a different operating point may not be valid and a new control-loop is to be estimated based on current operating point. However, linearizing the complex power system model every time as the operating point changes are tedious and an accurate model of the power system for the current operating point may not be always readily available. 

To overcome the drawbacks of linearization based methods, measurement based methods are adapted. In such methods, using the measurements, the MIMO model of the power system can be estimated directly without the need for having a detailed model of the power system. Since the changing dynamics/operating points of the system are captured in the measurement data, the MIMO model thus obtained can be used for mode and control loop estimation. The measurement-based methodologies gained interest with the advent of phasor measurement unit (PMUs), and modern communication infrastructure. A mode estimation method based on prony analysis is discussed  in \cite{refc1}, however the mutual coupling between different MIMO loops is not considered in the mode estimation. In \cite{refc2}, a measurement based method is discussed which is based on using simplified power system model and wide-area measurement system (WAMS) data. However the problems related to the selection of optimal control loop is not addressed here. The use of PMU data for wide-area control of STATCOM is addressed in \cite{refc3}, but the effect of mutual coupling during mode estimation and optimal signal selection is not represented. A power system model using MIMO identifcation based on auto-regressive moving average exogenous (ARMAX) methodology is developed in \cite{ref112} but an optimal method for signal selection is not addressed.

To address some of the existing issues, in this paper, a novel measurement based MIMO system identification technique for wide-area control considering mutual coupling between different control loops is presented. The identified MIMO system is used for 1) inter-area oscillation frequency estimation, 2) optimal control loop estimation, and 3) design of wide area controller.
To summarize, the major contributions of this work are:
\begin{itemize}
\setlength{\itemindent}{0em}
  \item A measurement based MIMO framework considering the mutual coupling of loops for mode estimation. 
  \item A method for online optimal control loop selection for WADC using residue analysis based on measurement data.
  \item An online WADC based on DLQR and KF techniques which uses the extracted state space matrices from identified MIMO system.
  \item A real-time co-simulation test bed for measurement based MIMO and WADC considering mutual coupling between different control loops.
\end{itemize}

The rest of the paper is organized as follows: Section II discusses the overall architecture and power system division based on coherency is discussed. In section III, measurement based MIMO identification is discussed and section IV discusses real-time implementation of MIMO identification. Section V discusses optimal wide area control loop selection and Section VI discusses measurement based DLQR and Kalman filter design. Section VII discusses simulation results and Section VIII concludes the paper.

\section{Dividing Power System Based on Coherency and Overall Architecture}

\begin{figure}[h]
\centering
\includegraphics[width=3.4 in,height=2.5in]{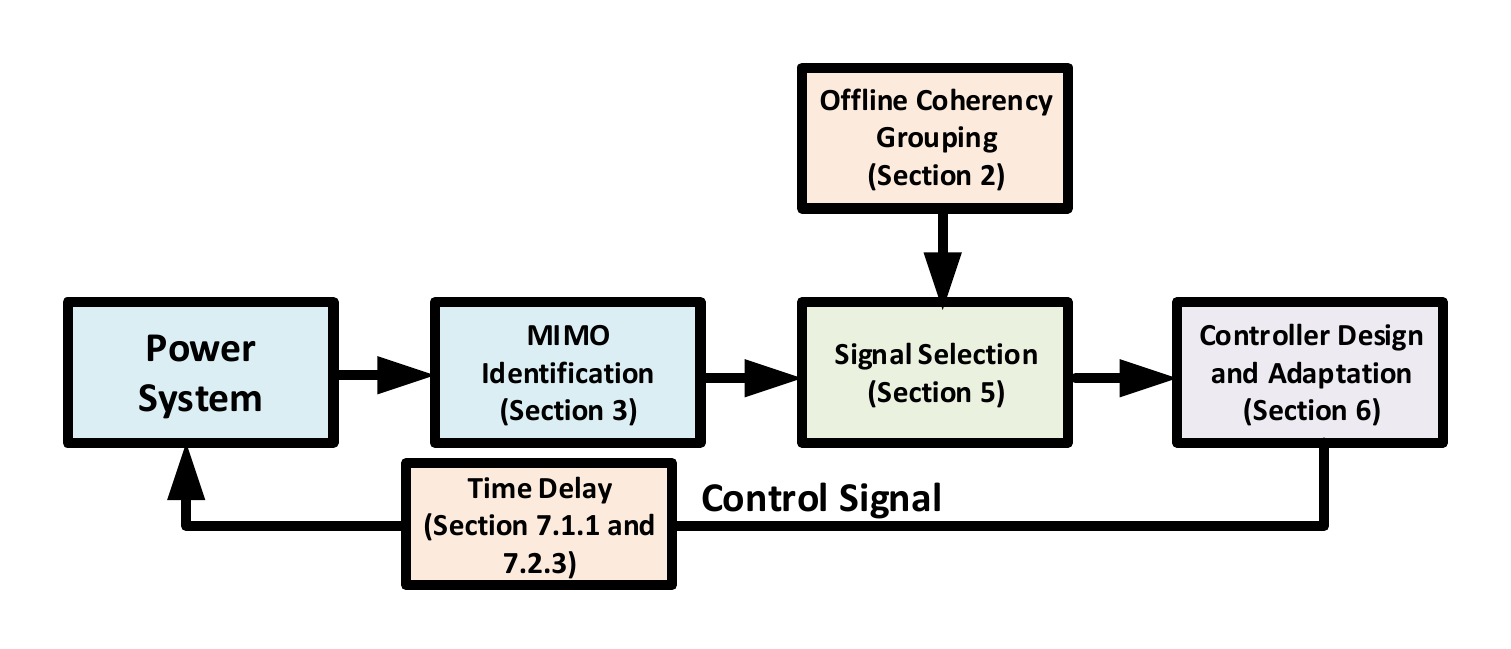}
\caption{Flow-chart of the proposed methodology}
\label{fig1aa}
\end{figure}

Design of the proposed architecture is based on dividing the power system into areas using coherency grouping and inter tie-line connections between areas. First, the active power flow deviation ($\Delta P = P_{rated}-P_{actual}$) through the tie-lines are used as input signal to the WADC \cite{refsom1}. Then MIMO identification is designed to estimate an optimal control loop to damp the inter-area mode oscillation. Finally, WADC is designed with a measurement based discrete linear quadratic regulator (DLQR) integrated in kalman filtering (KF) technique. Fig. \ref{fig1aa} shows the flowchart of the proposed methodology. The efficacy of the proposed algorithm is verified by implementing on a RTDS/RSCAD  and MATLAB   real-time test bed using two-area and IEEE 39 bus power system models. 

For grouping of power grid, slow coherency grouping method is considered. Please refer \cite{refa5} for further details of slow coherency grouping method. This algorithm is implemented on two-area power system (Fig. \ref{fig1a}) as well as IEEE 39 bus (Fig. \ref{fig6a}) power system models. Further, tie-line power flow information is used in deciding the wide-area control action.
\subsection{Coherency grouping of two-area system model}
\begin{figure}[h]
\centering
\includegraphics[width=3.5in,height=2in]{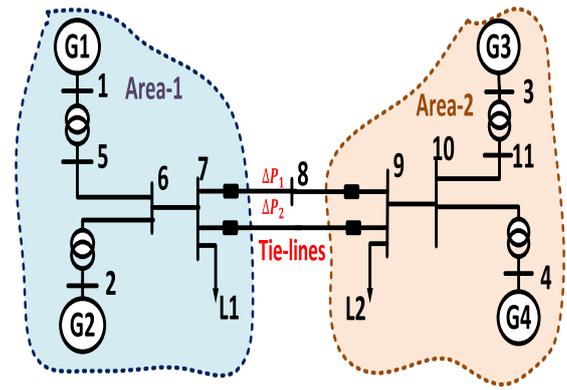}
\caption{Two-area power system model}
\label{fig1a}
\end{figure}

\begin{figure}[h]
\centering
\includegraphics[width=3.5in]{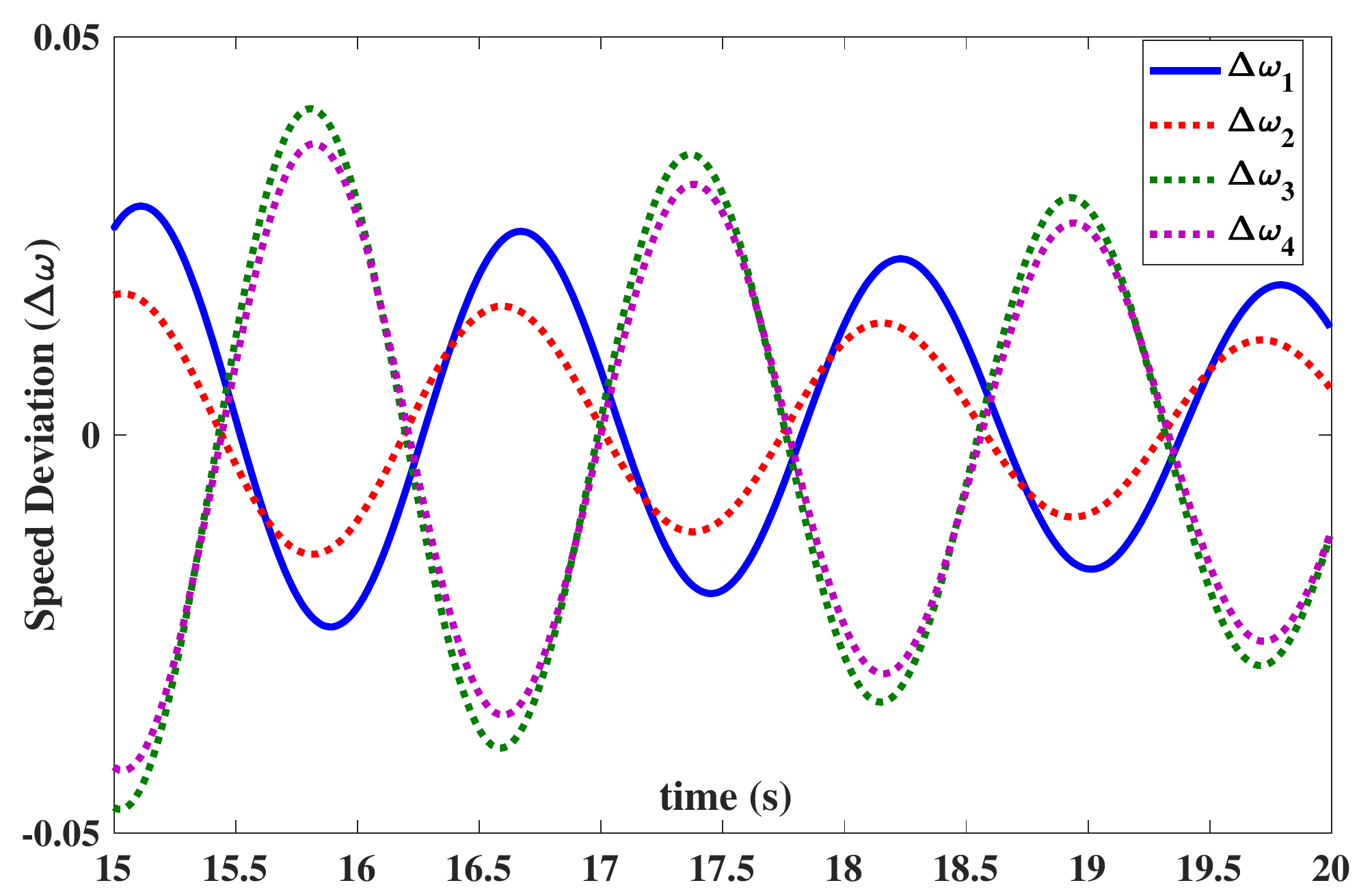}
\caption{Speed deviation (rad/s) (two-area system)}
\label{fig2a}
\end{figure}

On applying the slow coherency grouping method  to the two-area power system model, it is found that generators 1 and 2 are in one area and the remaining generators are in other area. To validate this grouping from the actual response of the system a 3-ph fault is created at 4s for a duration of 0.05s on bus-8 and generators inter-area dynamics are observed. Fig. \ref{fig2a} shows the speed deviation of generators from its rated speed (i.e $2\pi f$, where $f$ is fundamental frequency in Hz). It can be seen that generators 1,2 swing against generators 3 and 4. Based on this grouping the test model is divided into two areas (groups) as shown in Fig. \ref{fig1a} and Table \ref{tab1}. From Fig. \ref{fig1a} it can be seen that lines connecting Bus-7 to Bus-8 to Bus-9, and Bus-7 to Bus-9 are the two tie-lines. The active power flow deviation of these two lines are represented as $\Delta P_1$ and $\Delta P_2$ respectively.

\begin{table}[!b]
\processtable{Coherency grouping of power system models\label{tab1}}
{\begin{tabular*}{20pc}{@{\extracolsep{\fill}}lll@{}}\toprule
 Test System & Slow Coherency\\
\midrule
Two Area & Group-1: 1,2 \\
     & Group-2: 3,4 \\
\hline\\
IEEE 39-BUS & Group-1: 4,5,6,7\\
     & Group-2: 1,8,9\\
     & Group-3: 2,3\\
     & Group-4: 10\\
\botrule
\end{tabular*}}{}
\end{table}

\subsection{Coherency grouping of IEEE 39 bus system}
\begin{figure}[!h]
\centering
\includegraphics[width=3.5in]{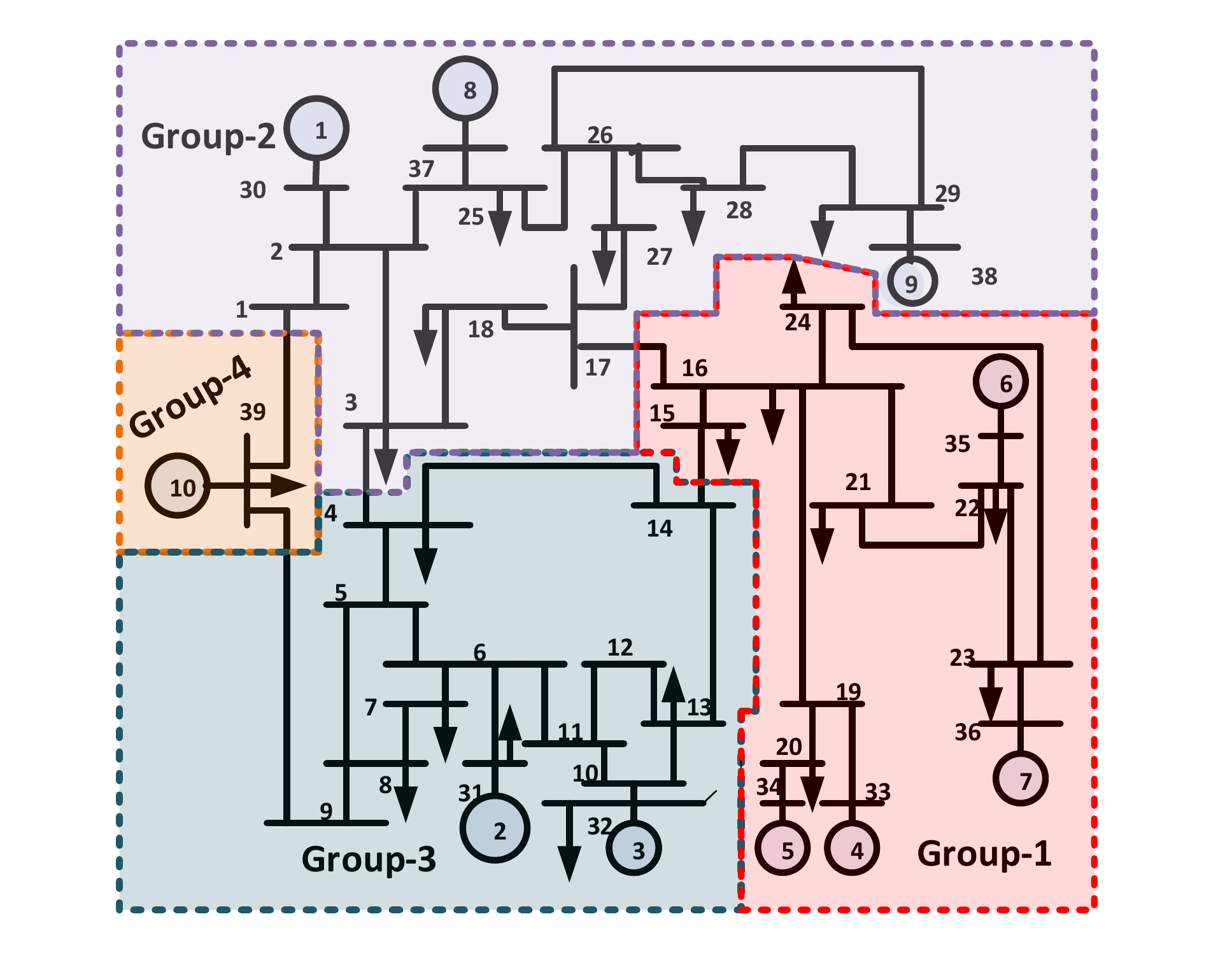}
\caption{39-bus power system model}
\label{fig3a}
\end{figure}

\begin{figure}[!h]
\centering
\includegraphics[width=3.5in]{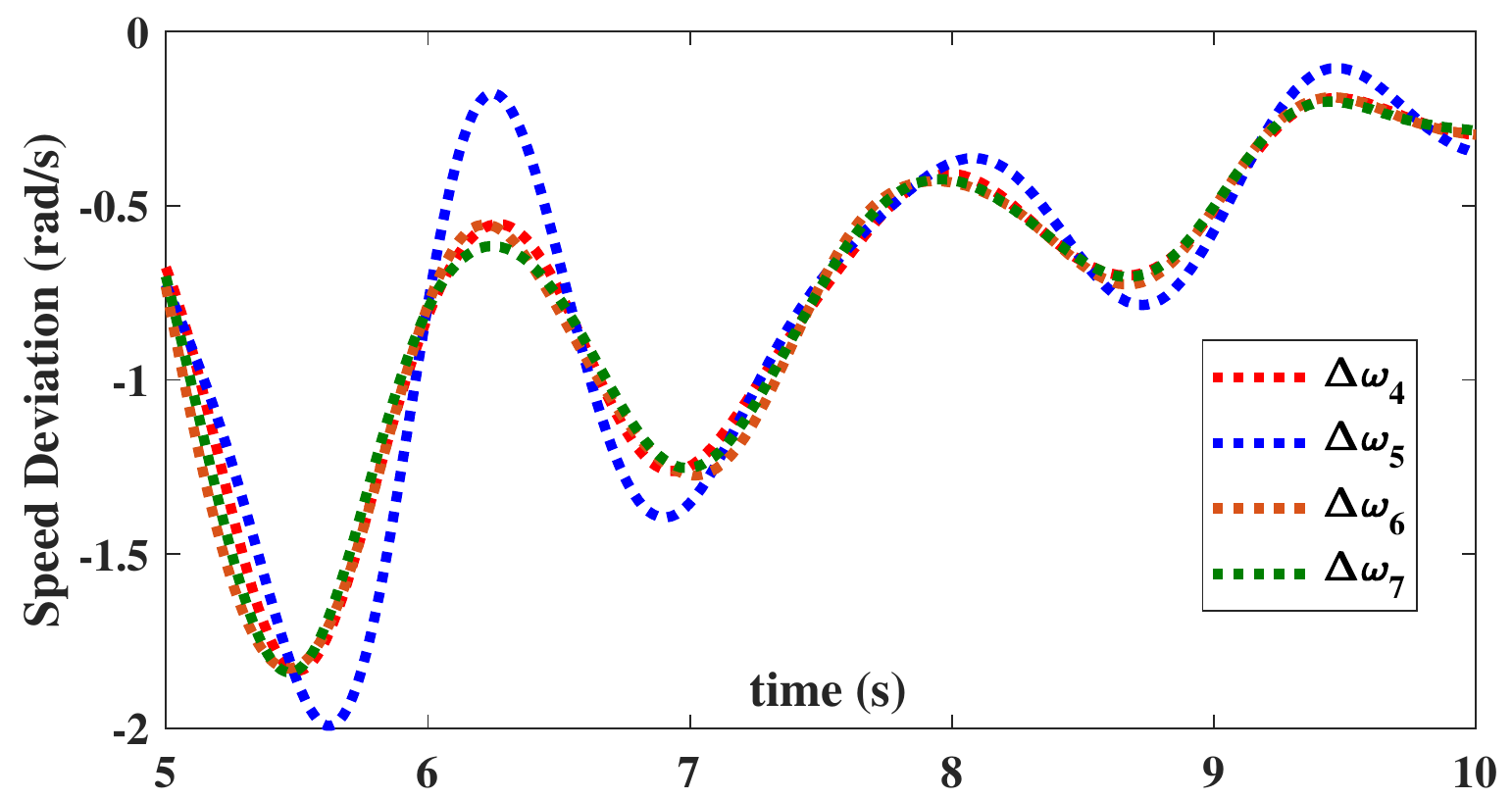}
\caption{Speed Deviation (Group-1)}
\label{fig4a}
\end{figure}

\begin{figure}[!h]
\centering
\includegraphics[width=3.5in]{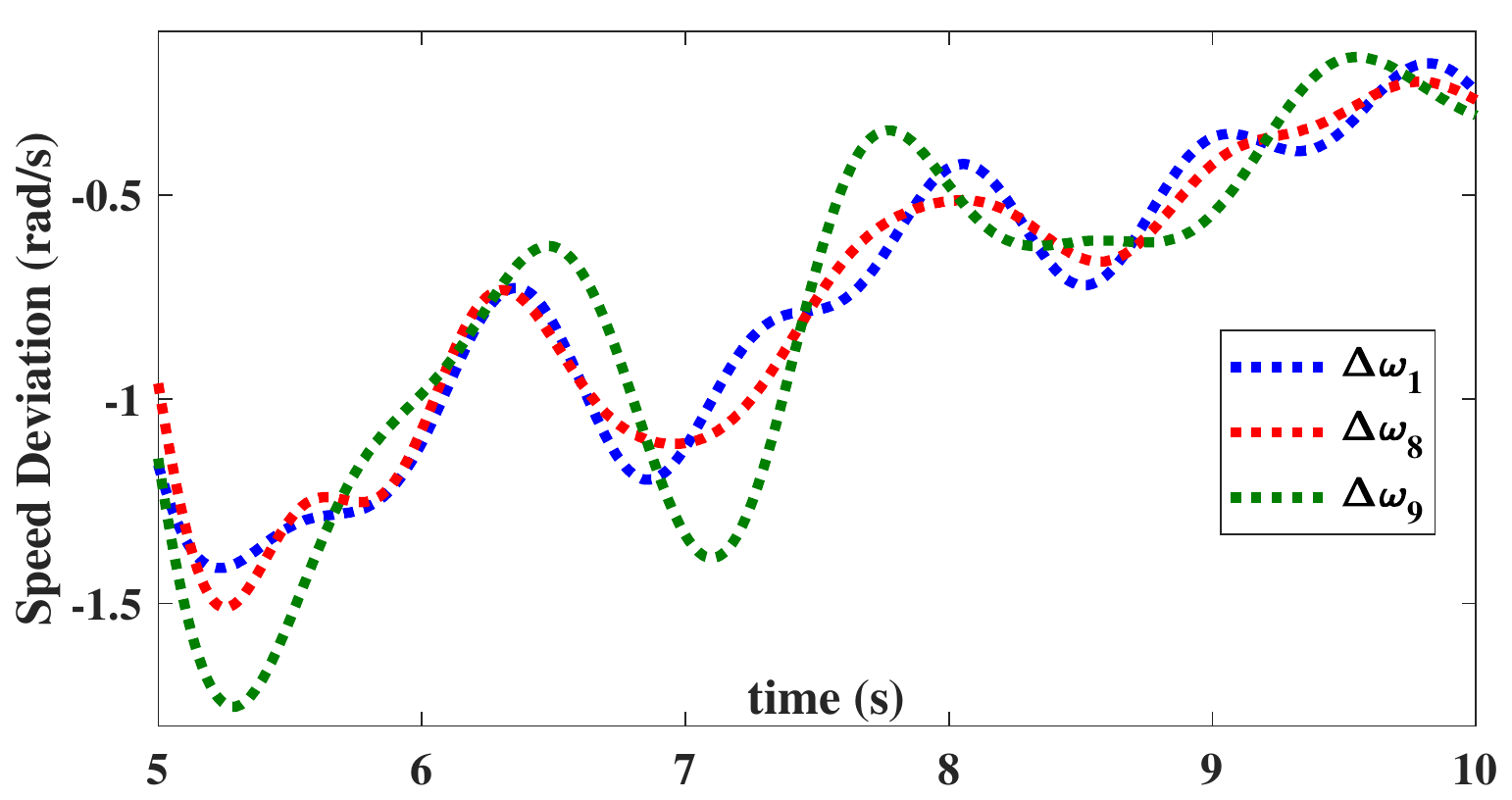}
\caption{Speed Deviation (Group-2)}
\label{fig5a}
\end{figure}

\begin{figure}[!h]
\centering
\includegraphics[width=3.5in]{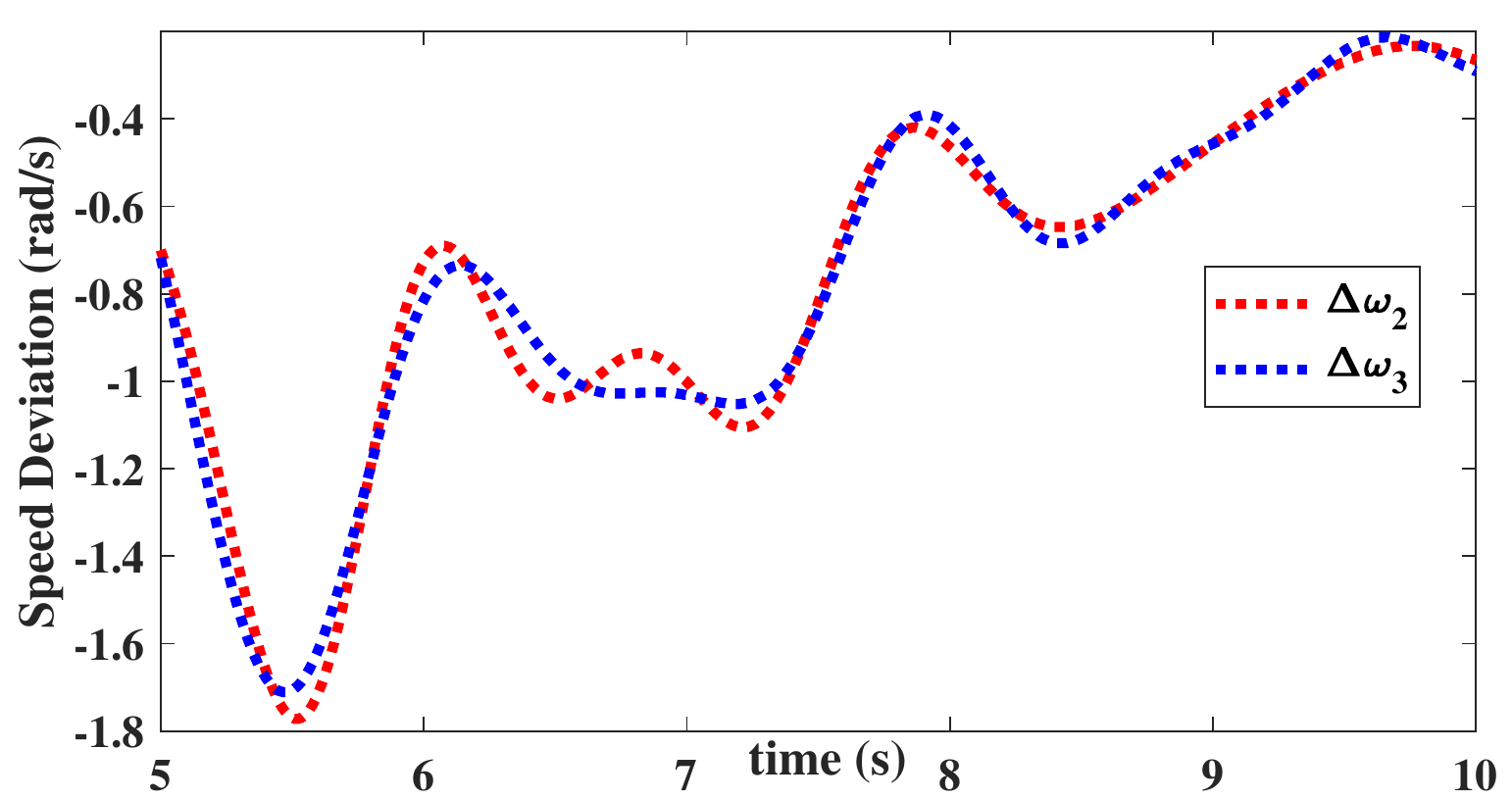}
\caption{Speed Deviation (Group-3)}
\label{fig6a}
\end{figure}

The IEEE 39 bus system (Fig. \ref{fig3a}) coherent groups of generators are obtained using the slow coherency grouping algorithm as shown in Table \ref{tab1}. For validation purpose, a 3-ph fault is created on Bus-14 at 4s for a duration of 0.1s. Fig. \ref{fig4a}, Fig. \ref{fig5a}, and Fig. \ref{fig6a} shows the speed deviations of generators in groups-1, 2, and 3 respectively. From Fig. \ref{fig3a} it can be seen that lines connecting Bus-39 to Bus-1, Bus-39 to Bus-9, Bus-3 to Bus-4, Bus-14 to Bus-15, and Bus-16 to Bus-17 are the tie-lines. The active power flow deviation of these tie-lines are represented as $\Delta P_1$, $\Delta P_2$, $\Delta P_3$, $\Delta P_4$, and $\Delta P_5$ respectively. 


\section{Measurement based MIMO Identification}
Most of the physical world systems are MIMO and traditionally the dynamics of these system are analyzed using linearized model at an operating point and estimating modes and mode shapes from the state space matrices thus obtained after linearization. However, linearizing a complex system requires accurate model which represents the current operating point of the system, This is often difficult to build. A measurement based MIMO identification can overcome modeling accuracy if identified  using measured data. However, the coupling effects between different input/output combinations should be considered while developing an MIMO model. The methodology is discussed using an example in Subsection \ref{3_1}.

\subsection{MIMO transfer function from state space matrices}\label{3_1}
Consider a system with two inputs, $u_1$ and $u_2$, and two outputs, $y_1$ and $y_2$ represented by differential equation \cite{ref12} as follows:
\begin{eqnarray}
\label{eqn1a}
y_1^{''}+a_1+y_1^{'}+a_0(y_1+y_2)=u_1(t)\\
\label{eqn2a}
y_2^{'}+a_2(y_2-y_1)=u_2(t)
\end{eqnarray}
where superscripts $y^{'}$ and $y^{''}$ indicates first and second order differential equations respectively. In state-space form \eqref{eqn1a} and \eqref{eqn2a} gives:
\begin{eqnarray}
\label{eqn3a}
x_1 = y_1\\
\label{eqn4a}
x_4 = y_2 = x_3^{'}\\
\label{eqn5a}
x_1^{'}=y_1^{'}=x_2\\
\label{eqn6a}
x_2^{'}=-a_1x_2-a_0(x_1+x_4)+u_1(t)\\
\label{eqn7a}
x_4^{'}=-a_2(x_4-x_1)+u_2(t)
\end{eqnarray}
Now rearranging \eqref{eqn3a} to \eqref{eqn7a} to get state space equations.
\begin{eqnarray}
     \begin{split}
    x^{'} &= \underbrace{\left[{\begin{array}{cccc}
   0 & 1 & 0 & 0\\
   -a_0 & -a_1 & 0 & -a_0\\
   0 & 0 & 0 & 1\\
   a_2 & 0 & 0 & -a_2\\
  \end{array} }\right]}_\text{\bf{A}}x +\underbrace{\left[{\begin{array}{cc}
   0 & 0\\
   1 & 0\\
   0 & 0\\
   0 & 1\\
  \end{array} }\right]}_\text{\bf{B}}\left[{\begin{array}{c}
   u_1\\
   u_2\\
  \end{array} }\right]\\
    \end{split}
     \label{eqn8a}
    \end{eqnarray}
\begin{eqnarray}
     \begin{split}
    \left[{\begin{array}{c}
   y_1\\
   y_2\\
  \end{array} }\right] &= \underbrace{\left[{\begin{array}{cccc}
   1 & 0 & 0 & 0\\
   0 & 0 & 0 & 1\\
  \end{array} }\right]}_\text{\bf{C}}x
    \end{split}
     \label{eqn9a}
    \end{eqnarray}
Converting \eqref{eqn8a} and \eqref{eqn9a} into transfer function form
\begin{eqnarray}
     G(s) &= \bf{C}(s\bf{I}-\bf{A})^{-1}\bf{B} &= \left[{\begin{array}{cc}
   G_{11}(s) & G_{12}(s)\\
   G_{21}(s) & G_{22}(s)\\
  \end{array} }\right]
  \label{eqn10a}
    \end{eqnarray}
where $\bf{I}$ is the identity matrix and
\begin{eqnarray*}
G_{11}(s)=\frac{s+a_2}{s^3+(a_1+a_2)s^2+(a_0+a_1a_2)s+2a_0a_2}\\
G_{12}(s)=\frac{-a_0}{s^3+(a_1+a_2)s^2+(a_0+a_1a_2)s+2a_0a_2}\\
G_{21}(s)=\frac{a_2}{s^3+(a_1+a_2)s^2+(a_0+a_1a_2)s+2a_0a_2}\\
G_{22}(s)=\frac{s^2+a_1s+a_0}{s^3+(a_1+a_2)s^2+(a_0+a_1a_2)s+2a_0a_2}\\
\end{eqnarray*}
Equation \eqref{eqn10a} can be represented in discrete time domain by replacing $s$ with $\frac{2(1-z^{-1})}{T_s(1+z^{-1})}$ (tustin approximation), where $T_s$ is the sampling time.
\begin{eqnarray}
     \begin{split}
    G(z^{-1}) &= \left[{\begin{array}{cc}
   G_{11}(z^{-1}) & G_{12}(z^{-1})\\
   G_{21}(z^{-1}) & G_{22}(z^{-1})\\
  \end{array} }\right]
    \end{split}
     \label{eqn11a}
    \end{eqnarray}

The objective here is to identify the MIMO transfer function as shown in \eqref{eqn10a} and \eqref{eqn11a} using the measurement data without having any knowledge of the physical system model. If MIMO identification is based on different input/output combinations separately, mutual coupling between loops are ignored. However, if we observe the individual transfer function of \eqref{eqn10a}, it can be seen that all the transfer functions have same denominator coefficients (i.e eigen values of the system) while numerator coefficients differ. This ensures mutual coupling between loops in MIMO system. So our objective is to formulate a MIMO identification using input/output measurement with same denominator coefficients but different numerator coefficients. 

\subsection{Measurement based MIMO identification considering mutual coupling}
\begin{figure}[!h]
\centering
\includegraphics[width=3.5in]{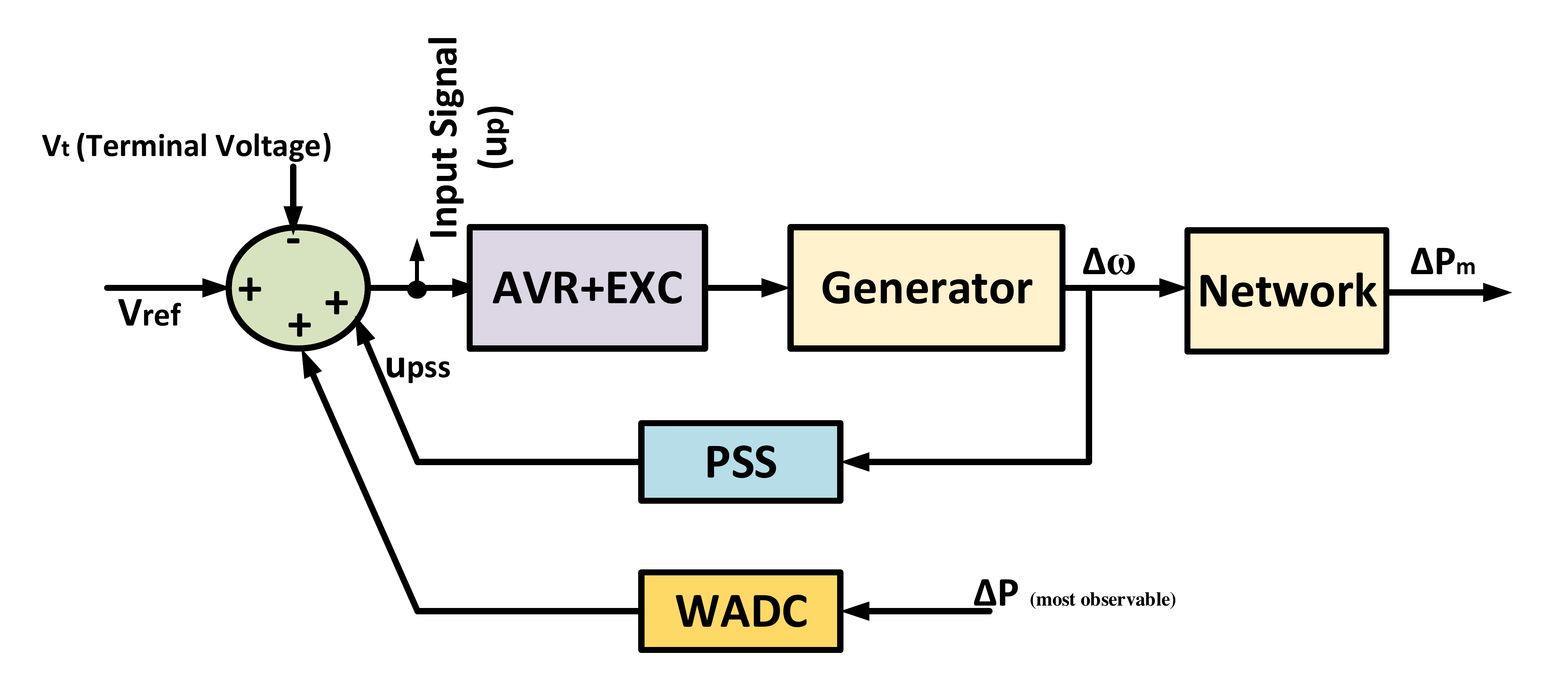}
\caption{Input and output measurements}
\label{fig6aa}
\end{figure}

For measurement based MIMO identification the input data $u_p$ is the voltage reference to the exciter of the synchronous machine and the output is the tie-line active power deviation, $\Delta P_{m}$. The input/output measurement channel is as illustrated in Fig. \ref{fig6aa}. The real-life system and EMT based real-time simulator is in discrete-domain, so the MIMO formulation of the power system is also modeled in discrete time domain with $p$ inputs and $m$ outputs as represented in \eqref{eqn12a}. The individual transfer function corresponding to each input/output loop is estimated using the proposed methodology for MIMO identification considering the mutual coupling  between different loops. This can be written as

\begin{eqnarray}
     \begin{split}
    \Delta P(z^{-1}) &=G(z^{-1})U(z^{-1})\\
    \end{split}
     \label{eqn12a}
    \end{eqnarray}
    where
 \begin{eqnarray*}
     \begin{split}
  \Delta P(z^{-1}) &= \left[{\begin{array}{c}
   \Delta P_{1}(z^{-1}) \\
   . \\
   . \\
   \Delta P_{m}(z^{-1}) 
  \end{array} }\right],
  U(z^{-1}) = \left[{\begin{array}{c}
   u_{1}(z^{-1}) \\
   . \\
   . \\
   u_{p}(z^{-1}) \\
  \end{array} }\right]\\
   G(z^{-1}) &=  \left[{\begin{array}{cccc}
   G_{11}(z^{-1}) &. &. &G_{1p}(z^{-1})\\
   .  &. &. &.\\
   .  &. &.  &.\\
   G_{m1}(z^{-1}) &. &. &G_{mp}(z^{-1})\\
  \end{array} }\right]
 \end{split} 
\end{eqnarray*}
and 
 \begin{eqnarray}
 \begin{split}
  G_{mp}(z^{-1}) &=\frac{\Delta P_{m}(z^{-1})}{u_{p}(z^{-1})} \\
                 &=\frac{b_0^h+b_1^hz^{-1}+...+b_k^hz^{-k}}{1+a_1z^{-1}+a_2z^{-2}+...+ a_kz^{-k}}
 \end{split}
  \label{eqn13a}
\end{eqnarray} 
where $u_p$ and $\Delta P_m$ are the input and output signal data,  $h$ is the element number in matrix, and $k$ is the order of transfer function. It can be seen from \eqref{eqn13a} that for all transfer functions the denominator coefficients are same but numerator coefficients differ. This ensures that mutual coupling effect is taken into account. The steps for MIMO transfer function identification is as follows:

\subsubsection{Step-1} 
The equation \eqref{eqn12a} is rewritten for $N$  observation window length as shown in \eqref{eqn14a}.
 
\begin{eqnarray}
  X_{His}^h = X_{Num}^h\left[{\begin{array}{c}
   b_0^h \\
   \vdots \\
   b_k^h \\
  \end{array} }\right] + X_{Den}^h\left[{\begin{array}{c}
   a_1 \\
   . \\
   . \\
   a_k \\
  \end{array} }\right]
  \label{eqn14a}
\end{eqnarray} 	
where 
\begin{eqnarray*}
 X_{His}^h = \left[{\begin{array}{c}
   \Delta P_{1}(z^{-1}) \\
   . \\
   . \\
   \Delta P_{m}(z^{-1}) 
  \end{array} }\right]
\end{eqnarray*} 	

\begin{eqnarray*}
  X_{Num}^h = \left[{\begin{array}{ccc}
   u_{p}(z-1) &. &u_{p}(z-k)\\
   .  &.  &.\\
   .  &.  &.\\
   u_{p}(z-N) &. &u_{p}(z-N+1-k)\\
  \end{array} }\right]
\end{eqnarray*} 	

\begin{eqnarray*}
  X_{Den}^h = \left[{\begin{array}{ccc}
   \Delta P_{m}(z-1) &. &\Delta P_{m}(z-k)\\
   .  &.  &.\\
   .  &.  &.\\
   \Delta P_{m}(z-N) &. &\Delta P_{m}(z-N+1-k)\\
  \end{array} }\right]
\end{eqnarray*}

 \subsubsection{Step-2} 
 For all input/output loops, concatenate $X_{His}^h$  and is represented as \eqref{eqn15a}

\begin{eqnarray}
 \underbrace{\left[{\begin{array}{c}
   X_{His}^1 \\
   X_{His}^2 \\
   . \\
   X_{His}^h\\
  \end{array} }\right]}_\text{\bf{$X_{H}$}} = \underbrace{\left[{\begin{array}{c}
   X_{Num}^1 \\
   X_{Num}^2 \\
   . \\
   X_{Num}^h\\
  \end{array} }\right]\left[{\begin{array}{c}
   b_0^1 \\
   \vdots \\
   b_k^1 \\
    b_0^2 \\
   \vdots \\
   b_k^2 \\
   \vdots \\
    b_0^h \\
   \vdots \\
   b_k^h \\
  \end{array} }\right]}_\text{\bf{$X_{N}$}}+ \underbrace{\left[{\begin{array}{c}
   X_{Den}^1 \\
   X_{Den}^2 \\
   . \\
   X_{Den}^h\\
  \end{array} }\right]\left[{\begin{array}{c}
   a_1 \\
   . \\
   . \\
   a_k \\
  \end{array} }\right]}_\text{\bf{$X_{D}$}}
    \label{eqn15a}
\end{eqnarray} 	

\subsubsection{Step-3} 
The numerator and denominator coefficients are estimated using an optimization algorithm where the optimization function is represented as \eqref{eqn16aa}
\begin{eqnarray}
      \label{eqn16aa}
 \text{minimize} \left\|X_{H}-\left[X_{N}+X_{D}\right]\right\|
\end{eqnarray}

Initial conditions for the numerator $(b_0^h,b_1^h... b_k^h)$ and denominator $(a_1,a_2... a_k)$ coefficients are estimated from a least squares technique formulated as constrained nonlinear multivariable function. For this the MATLAB  based $fmincon$ function is used to solve the unknowns. 




\section{Real-time implementation of MIMO identification}
\begin{figure}[h]
\centering
\includegraphics[width=3.5in]{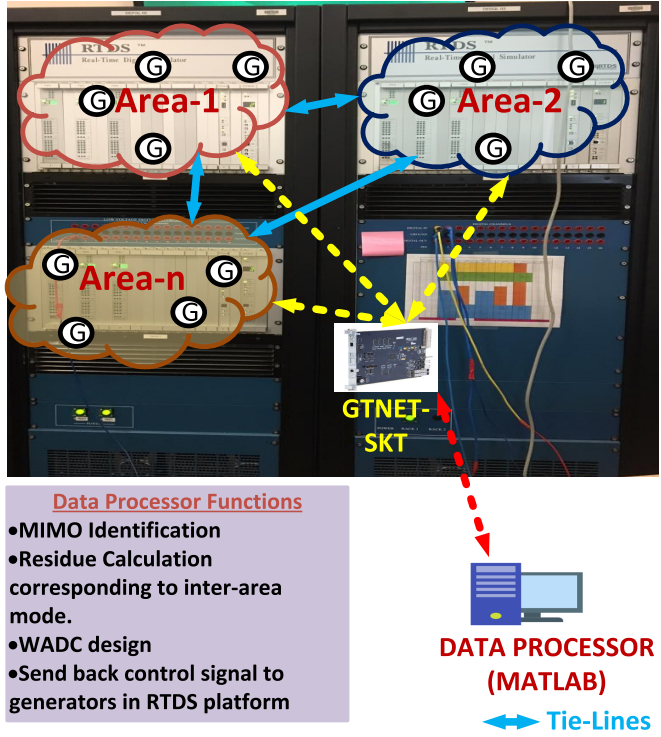}
\caption{Real-time experimental setup}
\label{fig7a}
\end{figure}

The proposed MIMO identification of the power system is implemented on a RTDS/RSCAD and MATLAB co-simulation test bed as shown in Fig. \ref{fig7a}. The power system is modeled in RTDS/RSCAD and the MIMO identification algorithm is implemented in MATLAB. The real-time measured data from the digital simulator is processed through an Ethernet based ocket connection \cite{ref13}. The data received is then processed for MIMO identification. Also the identified transfer function is used to estimate an optimal control loop. The control signel is then send back to the power system model in the digital simulator. Since this approach is implemented using RTDS, the process exactly emulates the real-life scenario. Thus the designed controller can be tested using this co-simulation platform before implementing in field.

\subsection{Data Criteria}

\begin{figure}[h]
\centering
\includegraphics[width=3.5in]{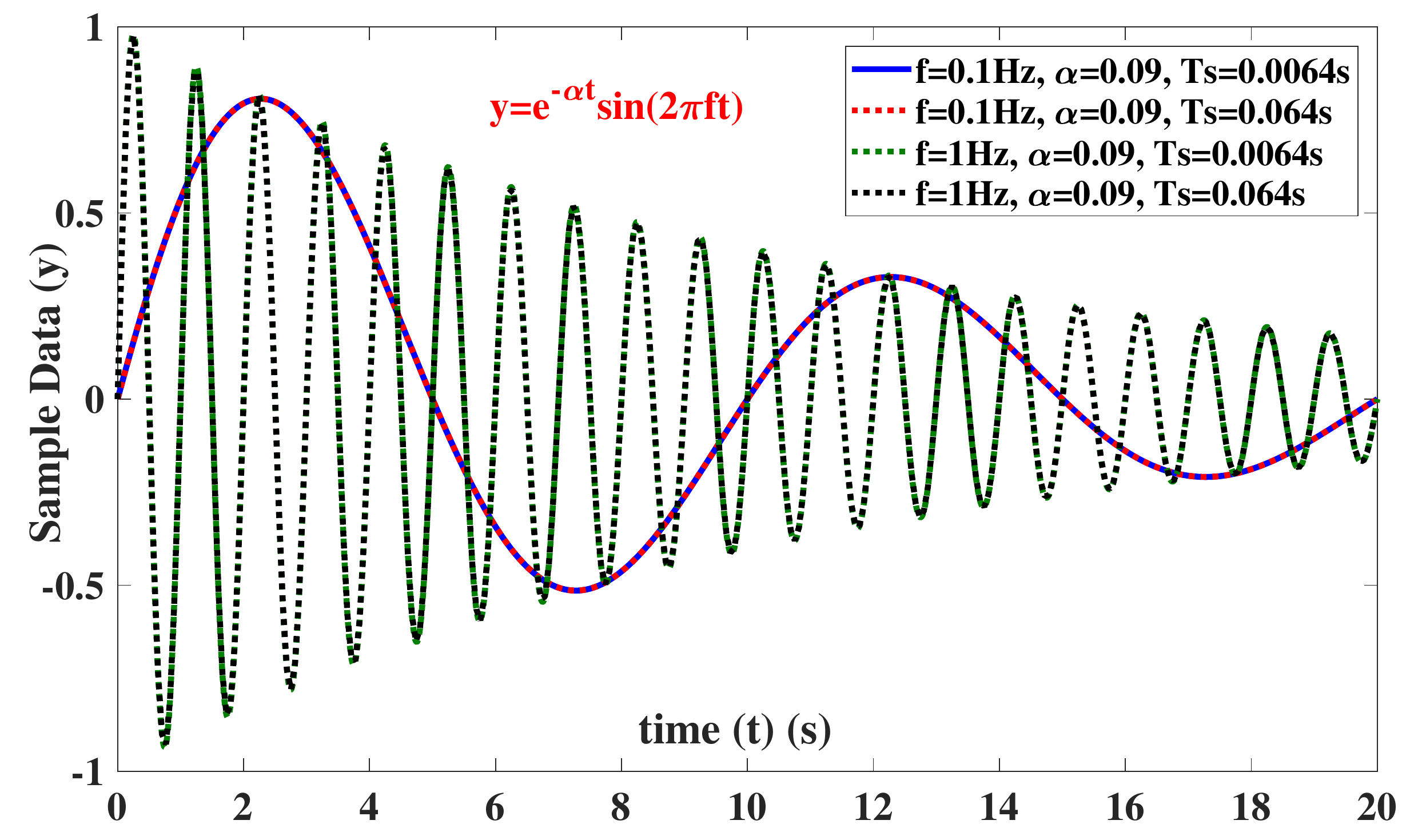}
\caption{Data requirement}
\label{fig7aa}
\end{figure}

The proposed algorithm is first validated on the two-area power system model (Fig. \ref{fig1a}). The measurement data $u_p$ and $\Delta P_m$ of all generators are extracted with a sampling time ($T_s$) of 0.0032s. The data-set is then down-sampled by a factor 10 so that the new sampling time will be 0.032s. The new sampling time after down-sampling is used to estimate inter-area modes. The inter-area frequency range is between 0.1 to 1Hz so for a successful estimation of inter-area frequency a window length of two cycles of the lowest frequency is considered as shown in Fig. \ref{fig7aa}.
\subsection{MIMO Identification two-area system}
\begin{figure}[h]
\centering
\includegraphics[width=3.5in]{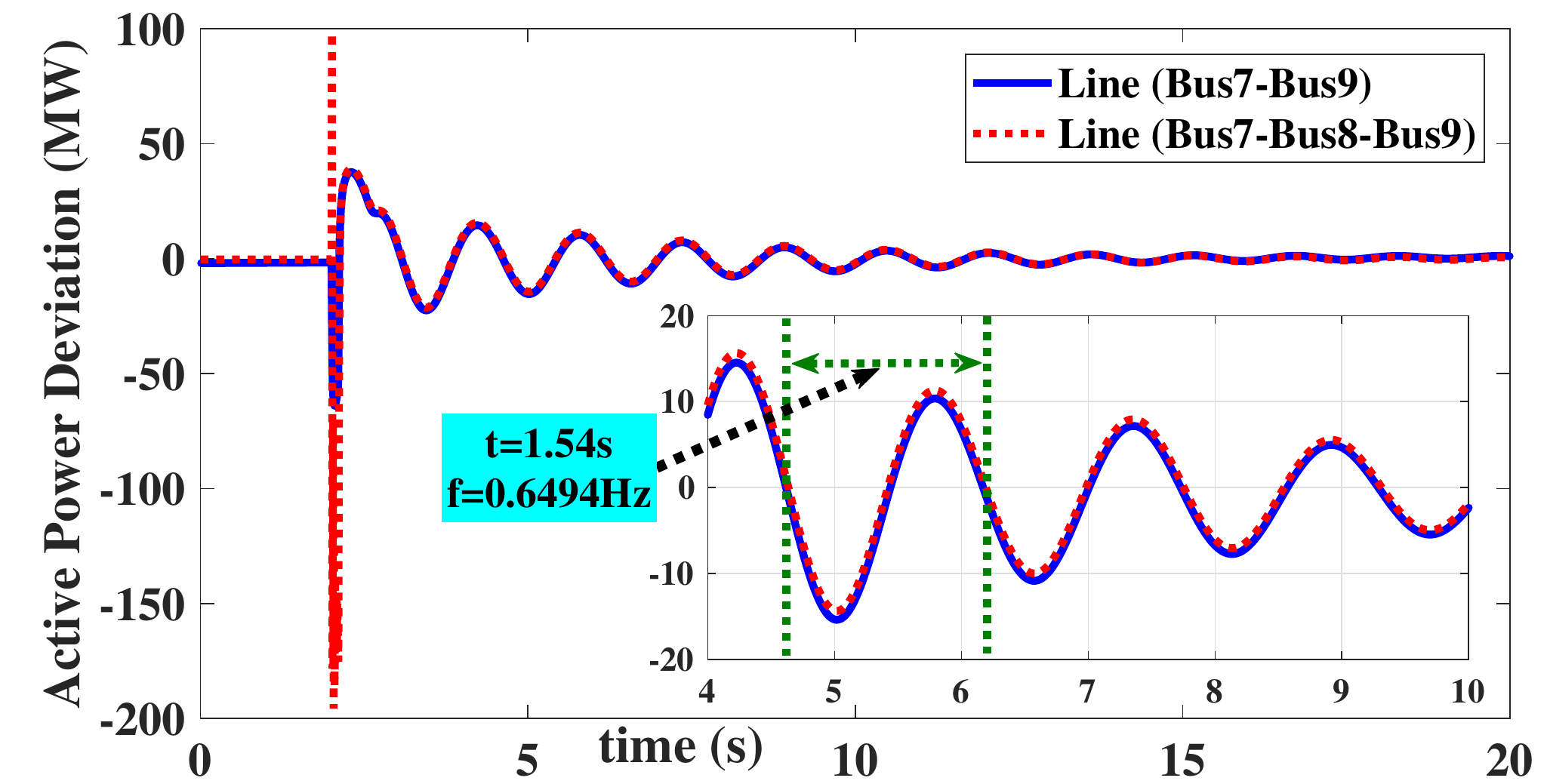}
\caption{Tie-lines active power deviation ($\Delta P_1$)}
\label{fig8a}
\end{figure}

\begin{figure}[h]
\centering
\includegraphics[width=3.5in]{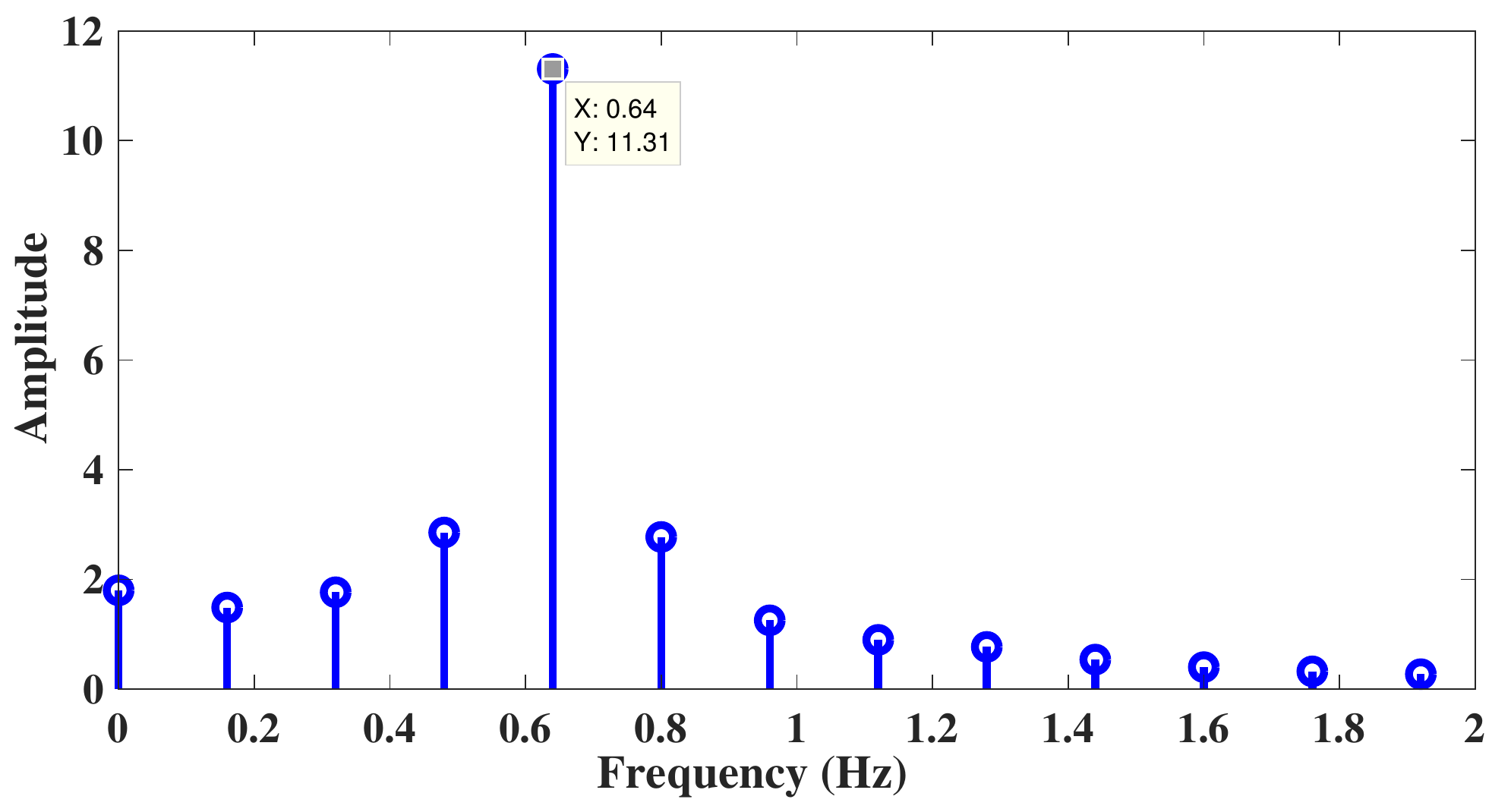}
\caption{FFT analysis of $\Delta P_1$}
\label{fig8a1}
\end{figure}

 To validate the effectiveness of the proposed MIMO identification algorithm in estimating the inter-area modes, a 3-ph fault is created on Bus-8 at 4s for a duration of 0.05s. The speed deviation after the fault is shown in Fig. \ref{fig8a}. From visual inspection, It can be seen that the approximate frequency of oscillation is 0.6494Hz. From fast fourier transform (FFT) algorithm on $\Delta P_1$ data, it can be seen that the frequency is 0.64Hz as well (see Fig. \ref{fig8a1}). The estimated transfer function between $\Delta P_1 \text{ (Bus-7 to Bus-8 to Bus-9)}$ and $u_4 \text{ (Generator-4)}$ is shown in \eqref{eqnx18a}. Using the transfer function, the frequency obtained from the proposed algorithm considering mutual coupling is 0.6362Hz. Table \ref{tab2} shows the inter-area mode comparison. 
 
\begin{align}
\begin{split}
   \bf{G_{14}}(z) &=\frac{\Delta P_{1}(z)}{u_{4}(z)} \\
                  &= \frac{59.2665 - 6.7181i}{1-( 0.9835 + 0.1257i)z^{-1}} \\
                  &+ \frac{59.2665 + 6.7181i}{1-(0.9835 - 0.1257i)z^{-1}} \\
                  & + \frac{-0.2852 +38.1373i}{1-(0.9300 + 0.2484i)z^{-1}} \\
                  &+ \frac{-0.2852 -38.1373i}{1-(0.9300 - 0.2484i)z^{-1}} \\
                  &+ \frac{-89.7368}{1-0.9961z^{-1}}-28.9939
\end{split}
  \label{eqnx18a}   
\end{align}

\begin{table}[h]
\processtable{Inter-area mode comparison\label{tab2}}
{\begin{tabular*}{20pc}{@{\extracolsep{\fill}}lll@{}}\toprule
 Test System & MIMO Identification & FFT\\
\midrule
Two Area & 0.6362Hz & 0.64Hz  \\
\hline\\
IEEE 39-BUS & Case-1: 0.6077Hz & Case-1: 0.599Hz\\
     & Case-2: 0.6098Hz & Case-2: 0.6024Hz\\
\botrule
\end{tabular*}}{}
\end{table}

\subsection{MIMO Identification of 39-bus system}
\begin{figure*}[h]
\centering
\includegraphics[width=7in]{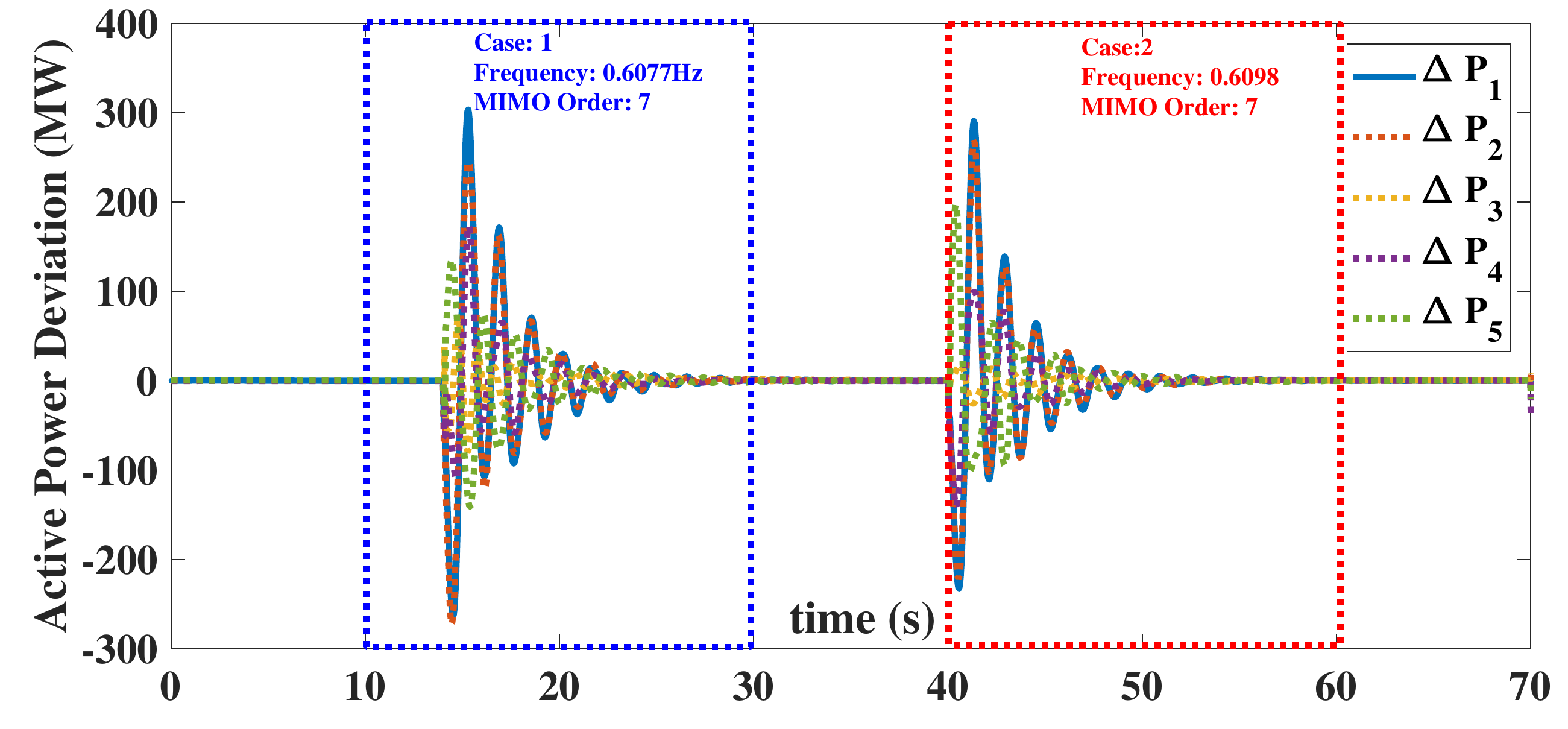}
\caption{39 bus tie-line active power deviation}
\label{fig10aaa}
\end{figure*}

In order to show the scalability and usefulness of the algorithm, case studies with multiple disturbance with multiple machines are evaluated using IEEE 39 bus system. Two sequential disturbance cases at different locations as shown in Fig. \ref{fig10aaa} are simulated.

\subsubsection{Case:1}\label{c1}
\begin{figure}[h]
\centering
\includegraphics[width=3.5in]{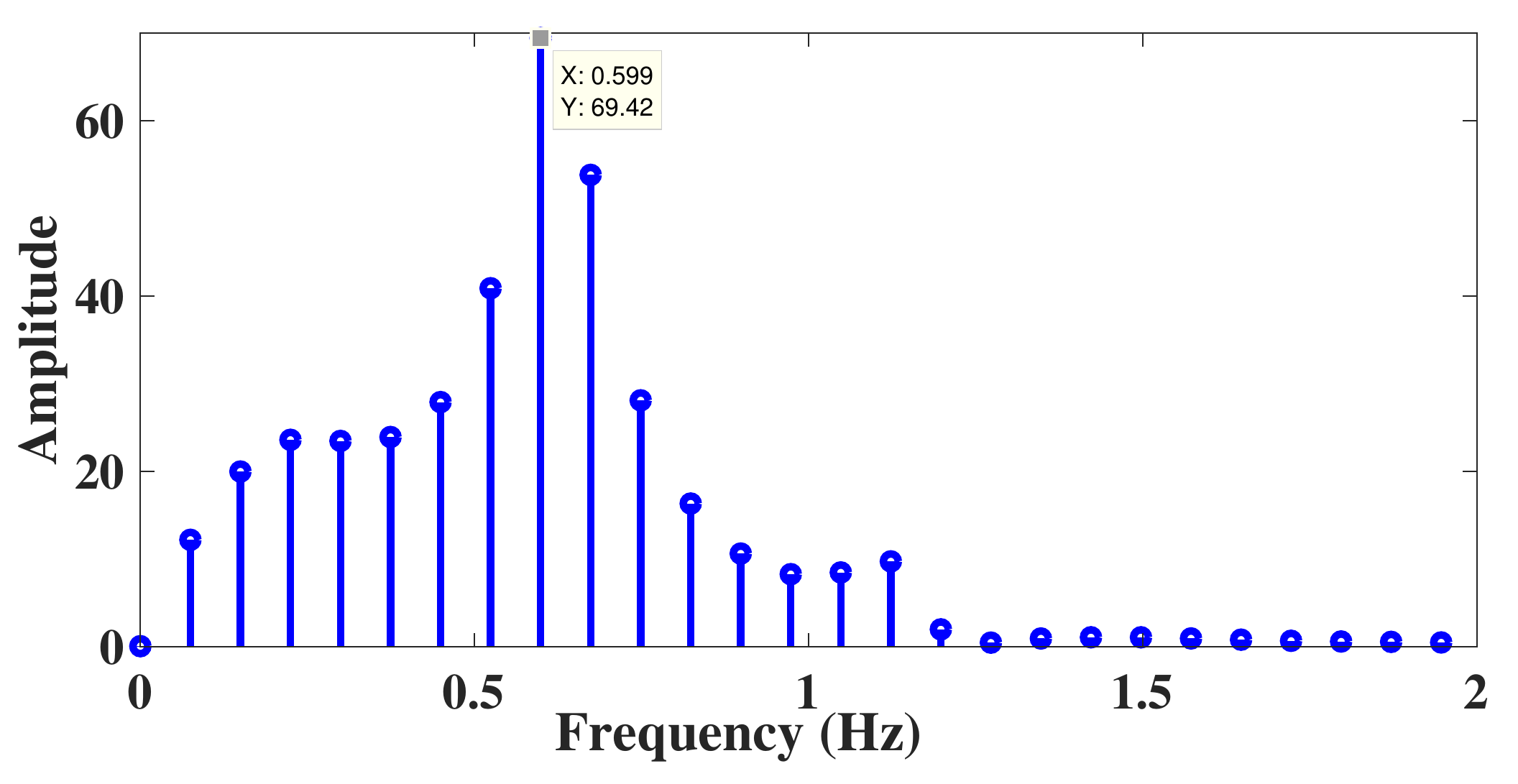}
\caption{FFT analysis of $\Delta P_1$ (Case:1)}
\label{fig10aa}
\end{figure}

In this case a 3-ph fault is created at 14s for a duration of 0.1s on bus-14. From Table \ref{tab2} it can be seen that, using the proposed method, estimated inter-area frequency is 0.6077Hz whereas from FFT analysis of $\Delta P_1$ (Bus-39 to Bus-1) data the inter-area frequency obtained is 0.599Hz (Fig. \ref{fig10aa}).

\subsubsection{Case:2}\label{c2}
\begin{figure}[h]
\centering
\includegraphics[width=3.5in]{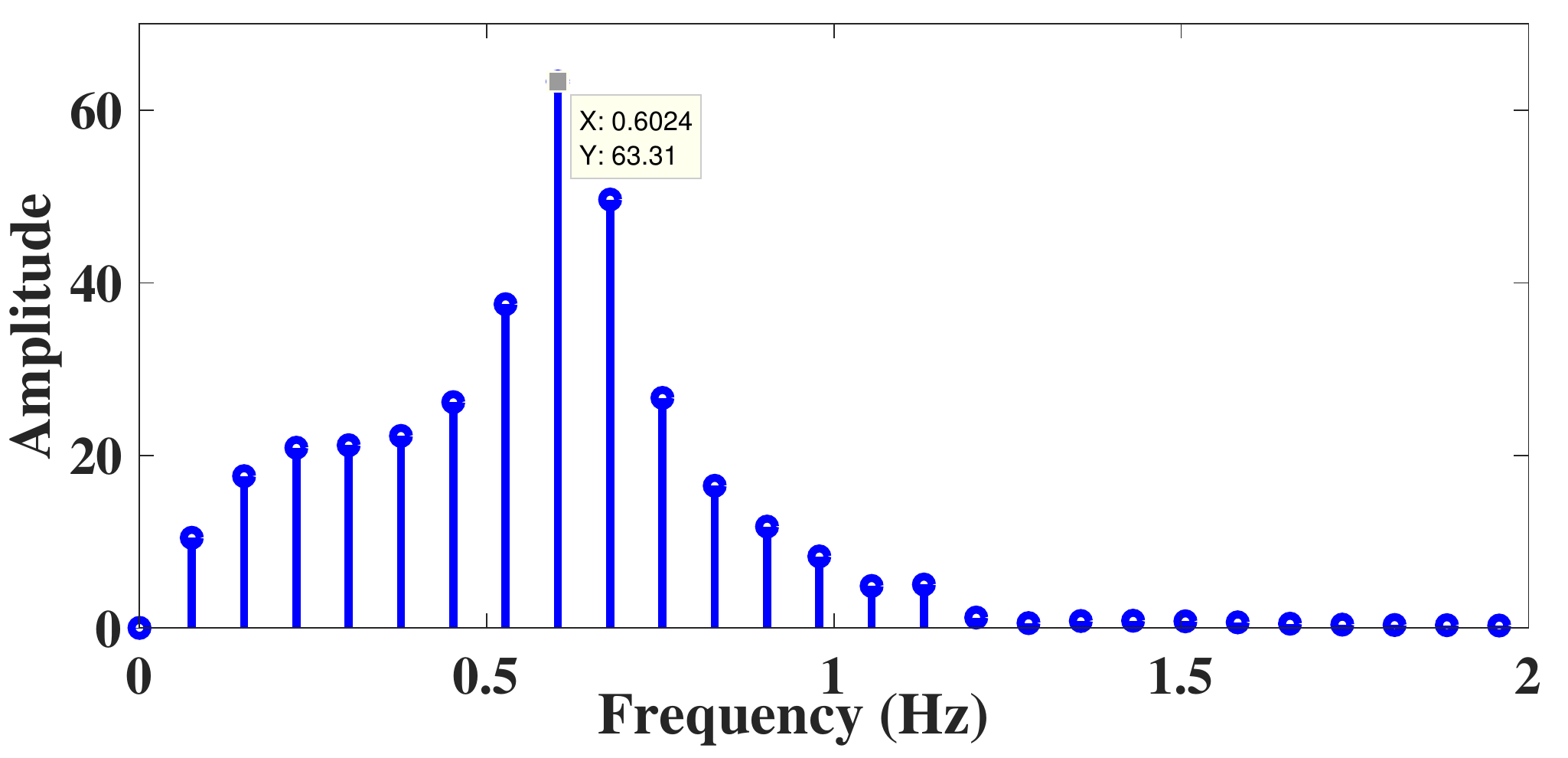}
\caption{FFT analysis of $\Delta P_1$ (Case:2)}
\label{fig11aa}
\end{figure}

In this case a 3-ph fault is created at 39.5s for a duration of 0.1s on bus-19. From Table \ref{tab2} it can be seen that from the proposed method the estimated inter-area frequency is 0.6098Hz whereas from FFT analysis of $\Delta P_1$ (Bus-39 to Bus-1) the inter-area frequency obtained is 0.6024Hz (Fig. \ref{fig11aa}). This proves that the proposed architecture can accurately represents the oscillation frequency.

\section{Optimal Wide Area Control Loop Selection}
After estimating the inter-area mode the objective is to damp that oscillation using WADC. However, in the MIMO power system model there are different input/output control loop combinations, it is important to estimate the most optimal control loop with which we can damp inter-area oscillations. This can be achieved by analyzing residues of the identified MIMO transfer functions corresponding to inter-area modes of interest \cite{refx11}. For this, \eqref{eqn13a} is converted to continuous time domain format and rewritten in partial fraction form as shown in \eqref{eqn18a}. 
\begin{align}
\begin{split}
   \bf{G_{mp}}(s) &=\frac{\Delta\omega_{m}(s)}{u_{p}(s)} \\
                  &= \frac{r_{mp}(1)}{s-p_1} + \frac{r_{mp}(2)}{s-p_2} + ..+ \frac{r_{mp}(j)}{s-p_j}+ k_{mp}(s) 
\end{split}
  \label{eqn18a}   
\end{align}

The transfer function between $\Delta P_1 \text{ (Bus-7 to Bus-8 to Bus-9)}$ and $u_4 \text{ (Generator-4)}$ using the proposed algorithm in continuous domain can be represented as \eqref{eqnx18a}.

\begin{align}
\begin{split}
   \bf{G_{14}}(z) &=\frac{\Delta P_{1}(s)}{u_{4}(s)} \\
                  &= \frac{-15.127 + 1211.8i}{s-( -1.2121 +     8.2017i)} \\
                  &+ \frac{-15.127 -     1211.8i}{s-(-1.2121 -     8.2017i)} \\
                  &    + \frac{1859.7 -     209.77i}{s-(-0.26881 +     3.9774i)} \\
                  &+ \frac{1859.7 +     209.77i}{s-(-0.26881 -     3.9774i)} \\
                  &+ \frac{-2804.3}{s-(-0.12207)}-10.142
\end{split}
  \label{eqnx18a}   
\end{align}
where $p$ and $r_{mp}$ are the poles and residues of the transfer function $G_{mp}(s)$. The residue ($r_{mp}(j)$) corresponding to inter-area mode ($p_j$) gives the information of the optimal control loop. For a mode $p_j$, the value of residue gives how controllable is the output $\Delta P_m$ and how observable is the input $u_p$, meaning how strong is the control loop to damp oscillations. It can be concluded that the control loop which has larger residue is the optimal loop.

\subsection{Identifying the optimal control loop for two-area system}
From Table \ref{tab2} it can be seen that the dominant inter-area mode for two-area power system model is 0.6362Hz. The residue analysis corresponding to the inter-area mode of frequency 0.6362Hz is shown in Table \ref{tab3}. From Table \ref{tab3} it can be seen that the maximum value of residue is for $\Delta P_1\rightarrow u_3$ loop which is the most optimal. From this, generator-3 is the most controllable and the active power deviation $\Delta P_1$ is the most observable signal.
\begin{table}[h]
\processtable{Optimal wide area control loop based on normalized residue analysis\label{tab3}}
{\begin{tabular*}{20pc}{@{\extracolsep{\fill}}lllll@{}}\toprule
output/input & $u_1$ & $u_2$ & $u_3$ & $u_4$\\
\midrule
$\Delta P_1\text{ (Bus-7 to Bus-8 to Bus-9)}$ & 0.3335 & 0.42875 & \bf{1} & 0.44531\\
\hline
$\Delta P_2\text{ (Bus-7 to Bus-9)}$ & 0.32973 & 0.42015 & 0.99597 & 0.47881\\
\botrule
\end{tabular*}}{}
\end{table}

\subsection{Identifying the optimal control loop for 39 bus system}
For 39 bus system, different cases are analyzed as discussed in Section \ref{c1} and \ref{c2} to illustrate how changes in operating conditions affect the inter-area modes and corresponding optimal control loops.
\subsubsection{Case: 1}
From Table \ref{tab2} dominant inter-area mode for 39 bus system, case-1 is 0.6077Hz. The residue analysis corresponding to inter-area mode of frequency 0.6077Hz for top three control loops are shown in Table \ref{tab4}. From Table \ref{tab4} it can be seen that generator-1, generator-8, and generator-9 are the most controllable and the corresponding observable signals are $\Delta P_2$, $\Delta P_1$, and $\Delta P_2$ respectively.
\begin{table}[h]
\processtable{Optimal control loop for IEEE 39-bus (Fault on Bus-14)\label{tab4}}
{\begin{tabular*}{20pc}{@{\extracolsep{\fill}}ll@{}}\toprule
Control Loop & Residue\\
\midrule
$\Delta\ P_2\text{ (Bus-39 to Bus-9)}$ to $u_1$ & 1.0 \\
\hline
$\Delta P_1\text{ (Bus-39 to Bus-1)}$ to $u_8$ & 0.8051 \\
\hline
$\Delta\ P_2\text{ (Bus-39 to Bus-9)}$ to $u_9$ & 0.7350 \\
\botrule
\end{tabular*}}{}
\end{table}
\subsubsection{Case: 2}
From Table \ref{tab2} dominant inter-area mode for 39 bus system, case-2 is 0.6098Hz. The residue analysis corresponding to inter-area mode of frequency 0.6098Hz for top three control loops are shown in Table \ref{tab5}. From Table \ref{tab5} it can be seen that generator-10, generator-1, and generator-8 are the most controllable and the corresponding observable signals are $\Delta P_5$, $\Delta P_5$, and $\Delta P_1$ respectively.
\begin{table}[h]
\processtable{Optimal control loop for IEEE 39-bus (Fault on Bus-19)\label{tab5}}
{\begin{tabular*}{20pc}{@{\extracolsep{\fill}}ll@{}}\toprule
Control Loop & Residue\\
\midrule
$\Delta\ P_5\text{ (Bus-16 to Bus-17)}$ to $u_{10}$ & 1.0 \\
\hline
$\Delta P_5\text{ (Bus-16 to Bus-17)}$ to $u_1$ & 0.6993 \\
\hline
$\Delta\ P_1\text{ (Bus-39 to Bus-1)}$ to $u_8$ & 0.4738 \\
\botrule
\end{tabular*}}{}
\end{table}

\begin{figure}[h]
\centering
\includegraphics[width=3.5in]{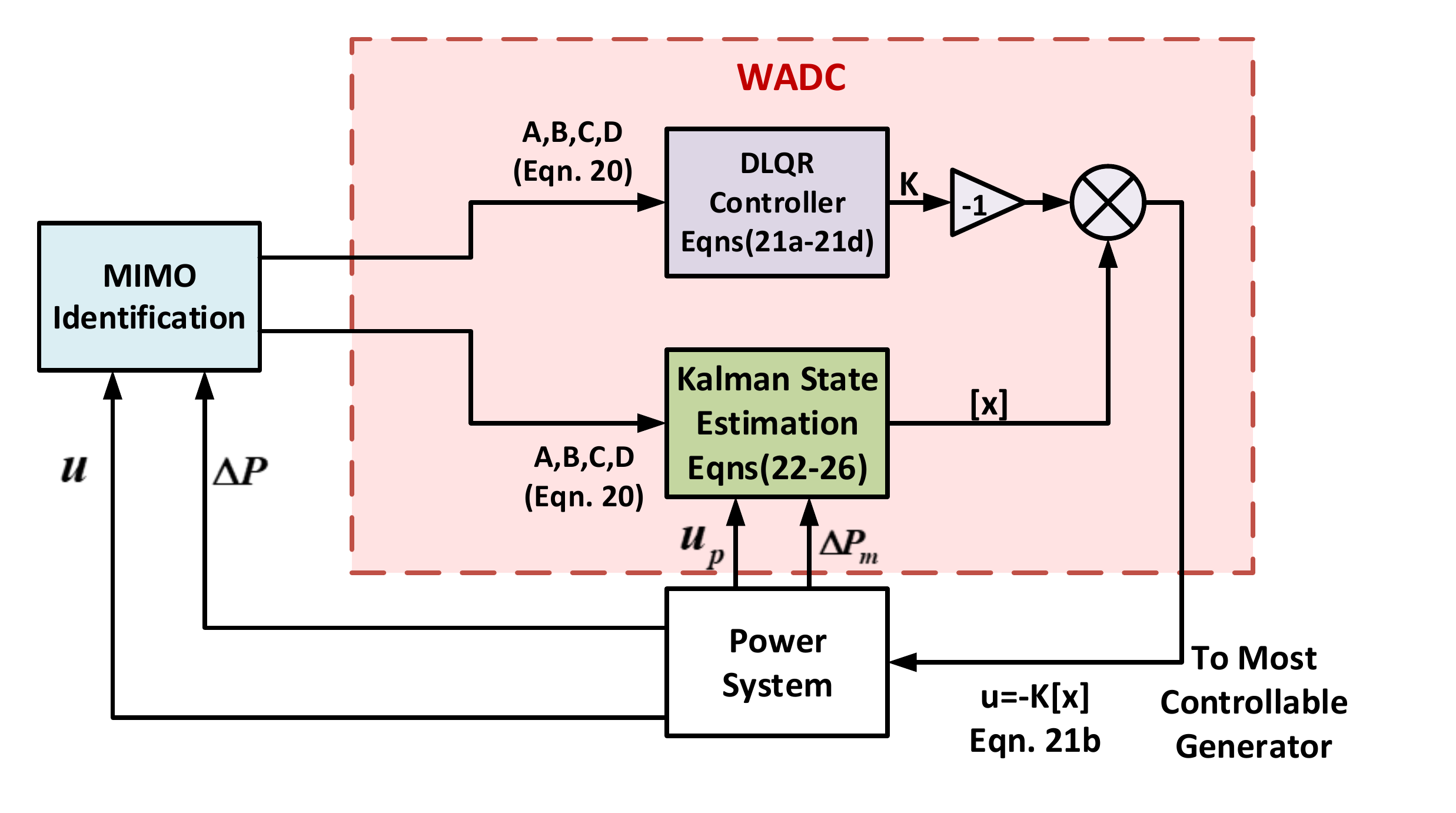}
\caption{WADC architecture}
\label{fig11bb}
\end{figure}

\section{Measurement based DLQR and Kalman filter design}
The WADC design is based on DLQR and KF. The WADC design architecture is as shown in Fig. \ref{fig11bb}.

\subsection{Discrete-time linear quadratic regulator}
The DLQR based controller uses state space matrices corresponding to optimal control loop for calculating the control gain. For this the discrete state space model is extracted from the MIMO transfer functions \eqref{eqn13a} and represented as follows:
\begin{align}
\begin{split}
   x_{k+1}&= A_kx_k+B_ku_k \\
              y_k &= C_kx_k+D_ku_k 
\end{split}
  \label{eqn19a}   
\end{align}
where $A$, $B$, $C$, and $D$ are the state, input, output, and feed forward matrices respectively.
The objective of the DLQR is to minimize the cost of the objective function represented as \eqref{eqn20a}, where $R = \rho I (\rho > 0)$, $Q = C_k^TC_k$ are the weight matrices and $N$ is the number of samples. The cost function can be minimized by calculating optimal control gain $K$ by solving \eqref{eqn20c}-\eqref{eqn20d}. The $P$ in \eqref{eqn20c} is the solution of the discrete algebraic Ricatti equations represented as \eqref{eqn20d}. The optimal control signal is represented as \eqref{eqn20b} \cite{ref15} which is sent to most controllable generator estimated from residue analysis.

\begin{subequations}
\begin{align}
   J &= \sum_{k=0}^{N} ({x_k^T}Qx_k + {u_k^T}Ru_k)  \label{eqn20a}\\  
   u_{k} &= -K_kx_k  \label{eqn20b}\\
   K_k &= (R + B_k^TP_{k+1}B_k)^{-1}B_k^T P_{k+1} A_k  \label{eqn20c}\\
   \begin{split}
 P_{k-1}&= Q +A_k^T P_k A_k \\ 
                &-A_k^T P_k B_k(R+B_k^T P_k B_k)^{-1} B_k^T P_k A_k        \label{eqn20d}
 \end{split}
\end{align}
\end{subequations}

\begin{table}[h]
\processtable{Estimated LQR gains ($K$) for two-area system\label{tab2aaa}}
{\begin{tabular*}{20pc}{@{\extracolsep{\fill}}ll@{}}\toprule
 Control Loop & Gain (K)\\
\midrule
$\Delta P_1$ to $u_3$ & [0.089446      -0.1349     0.073803    -0.033455     0.019523]  \\
\botrule
\end{tabular*}}{}
\end{table}

\begin{table}[h]
\processtable{Estimated LQR gains ($K$) for  IEEE 39 bus system (Case:1)\label{tab2aab}}
{\begin{tabular*}{20pc}{@{\extracolsep{\fill}}ll@{}}\toprule
 Control Loop & Gain (K)\\
\midrule
$\Delta P_2$ to $u_1$ & [0.0041   -0.0065   -0.0029    0.0295   -0.0407    0.0233   -0.0190]  \\
\hline
$\Delta P_1$ to $u_8$ & [-0.0467 0.1225 -0.2635 0.2954 -0.1783 0.0519 -0.0188]  \\
\hline
$\Delta P_2$ to $u_9$ & [-0.0296    0.0884   -0.2031    0.2372   -0.1477    0.0448   -0.0188]  \\
\botrule
\end{tabular*}}{}
\end{table}

\begin{table}[h]
\processtable{Estimated LQR gains ($K$) for  IEEE 39 bus system (Case:2)\label{tab2aac}}
{\begin{tabular*}{20pc}{@{\extracolsep{\fill}}ll@{}}\toprule
 Control Loop & Gain (K)\\
\midrule
$\Delta P_5$ to $u_{10}$ & [-0.0919    0.3275   -0.4145    0.2566   -0.1636    0.0935   -0.0207]  \\
\hline
$\Delta P_5$ to $u_1$ & [-0.0858    0.3297   -0.4227    0.2606   -0.1645    0.0931   -0.0207]  \\
\hline
$\Delta P_1$ to $u_8$ & [-0.0445    0.1546   -0.1984    0.1249   -0.0807    0.0465   -0.0104]  \\
\botrule
\end{tabular*}}{}
\end{table}

Table \ref{tab2aaa} shows the estimated optimal control gains for two-area system. Table \ref{tab2aab} and Table \ref{tab2aac} shows the optimal control gains for 39-bus system Case 1 and 2 respectively.

\subsection{Kalman filtering based state estimation}
The optimal control signal in \eqref{eqn20b} is the product of the optimal control gain and states of the system corresponding to optimal control loop. Here, KF technique is adapted to estimate the states of the system. The state space model as shown in \eqref{eqn19a} is used to estimate the states. The predictor step and co-variance calculation is given by \eqref{eqn21a} and \eqref{eqn22a} respectively. 

\begin{eqnarray}
   x_{k+1} = A_kx_k + B_ku_k
  \label{eqn21a}\\
     L_k = A_kL_{k-1}A_k^T + Q_k
  \label{eqn22a}  
\end{eqnarray} 
where $Q$ and $L$ are co-variance of noise and state vector estimate respectively. The Kalman gain factor ($G$) is estimated as shown in \eqref{eqn23a}
 
\begin{eqnarray}
   G_k = L_kH_k^T(H_kL_kH_k^T+R_k)^{-1}
  \label{eqn23a}  
\end{eqnarray}	
where $R$ and $H$ are co-variance of measurement noise and and the observation matrix respectively. The corrector step is represented as shown in \eqref{eqn24a} and \eqref{eqn25a}.
 
\begin{eqnarray}
   x_k = x_{k+1} + G_k(z-Hx{k+1}) 
  \label{eqn24a}\\
     L_k = L_{k+1}- K_kH_kL_{k+1}
  \label{eqn25a}  
\end{eqnarray}	
Further details of kalman filtering based state estimation are reported in \cite{ref14}.

\section{Simulation Results}
The output of WADC is sent to the most controllable generator as shown in Fig. \ref{fig11bb}. Implementation results on two-area and IEEE 39 bus system models considering various disturbances are illustrated next.

\subsection{Simulation results of two-area system}
\begin{figure}[h]
\centering
\includegraphics[width=3.5in]{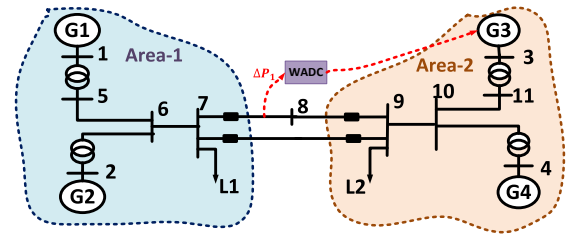}
\caption{Optimal control loop for two-area system}
\label{figs1}
\end{figure}

\begin{figure}[h]
\centering
\includegraphics[width=3.5in]{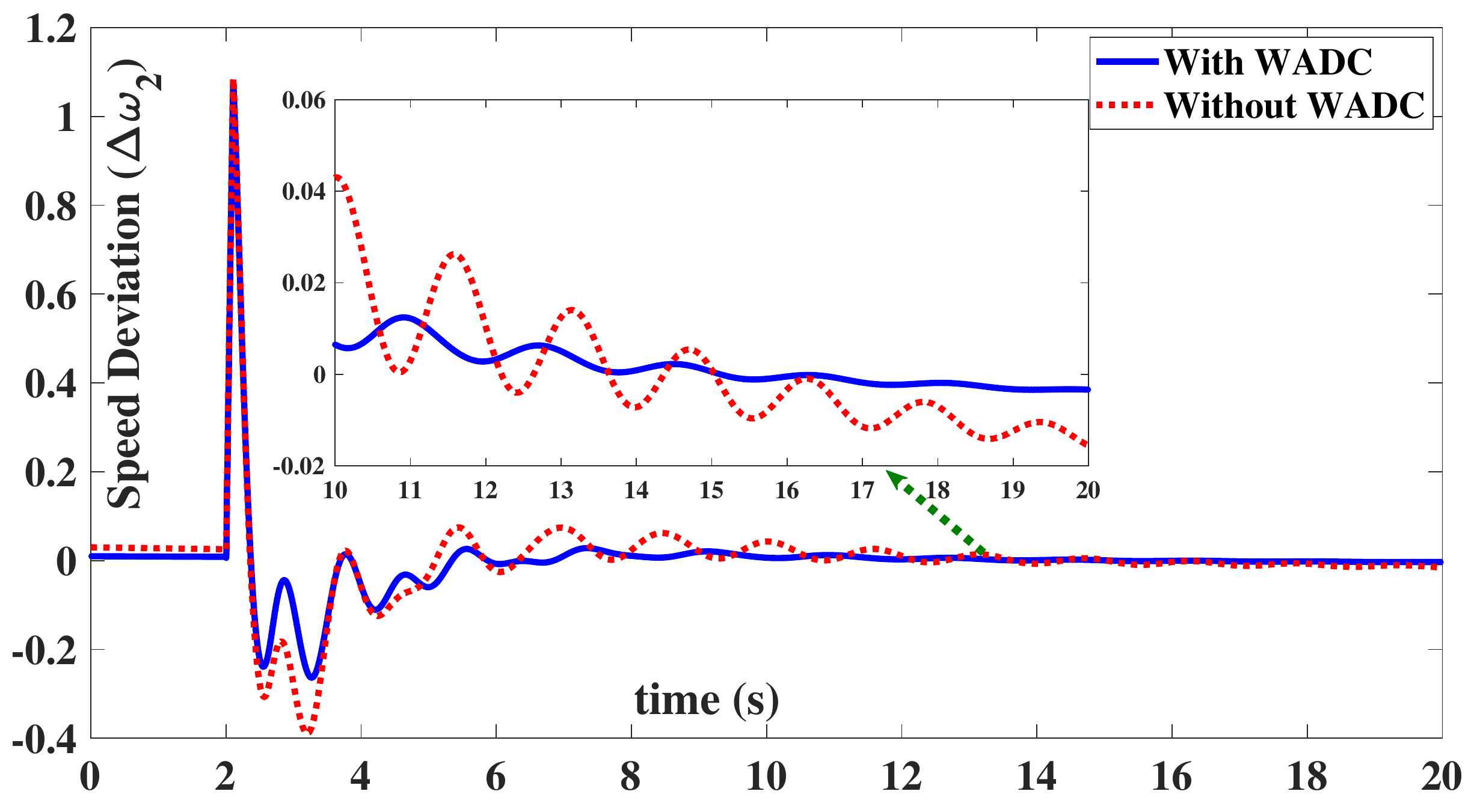}
\caption{Generator-2 speed deviation (rad/s)}
\label{figs2}
\end{figure}

\begin{figure}[h]
\centering
\includegraphics[width=3.5in]{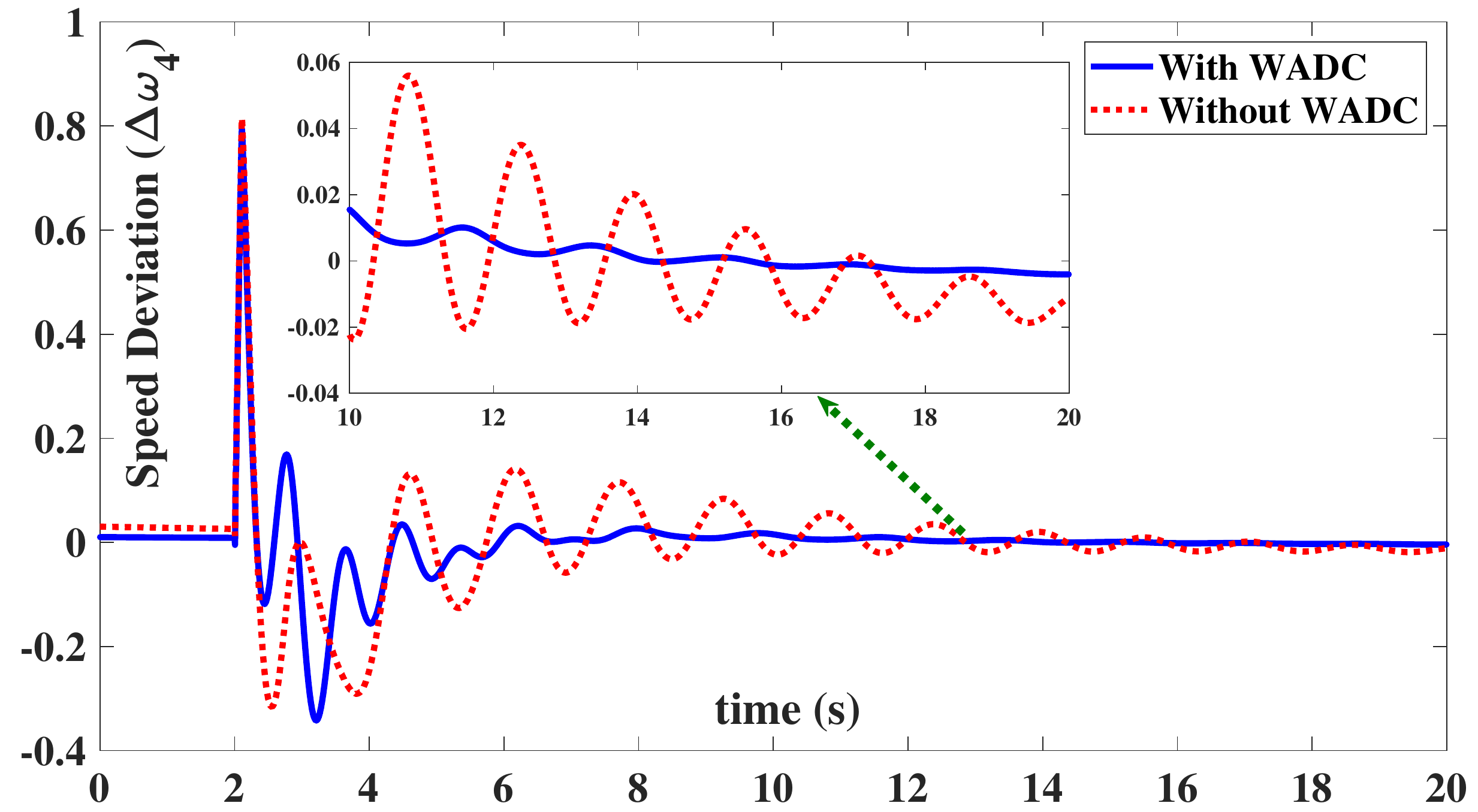}
\caption{Generator-4 speed deviation (rad/s)}
\label{figs3}
\end{figure}

\begin{figure}[h]
\centering
\includegraphics[width=3.5in]{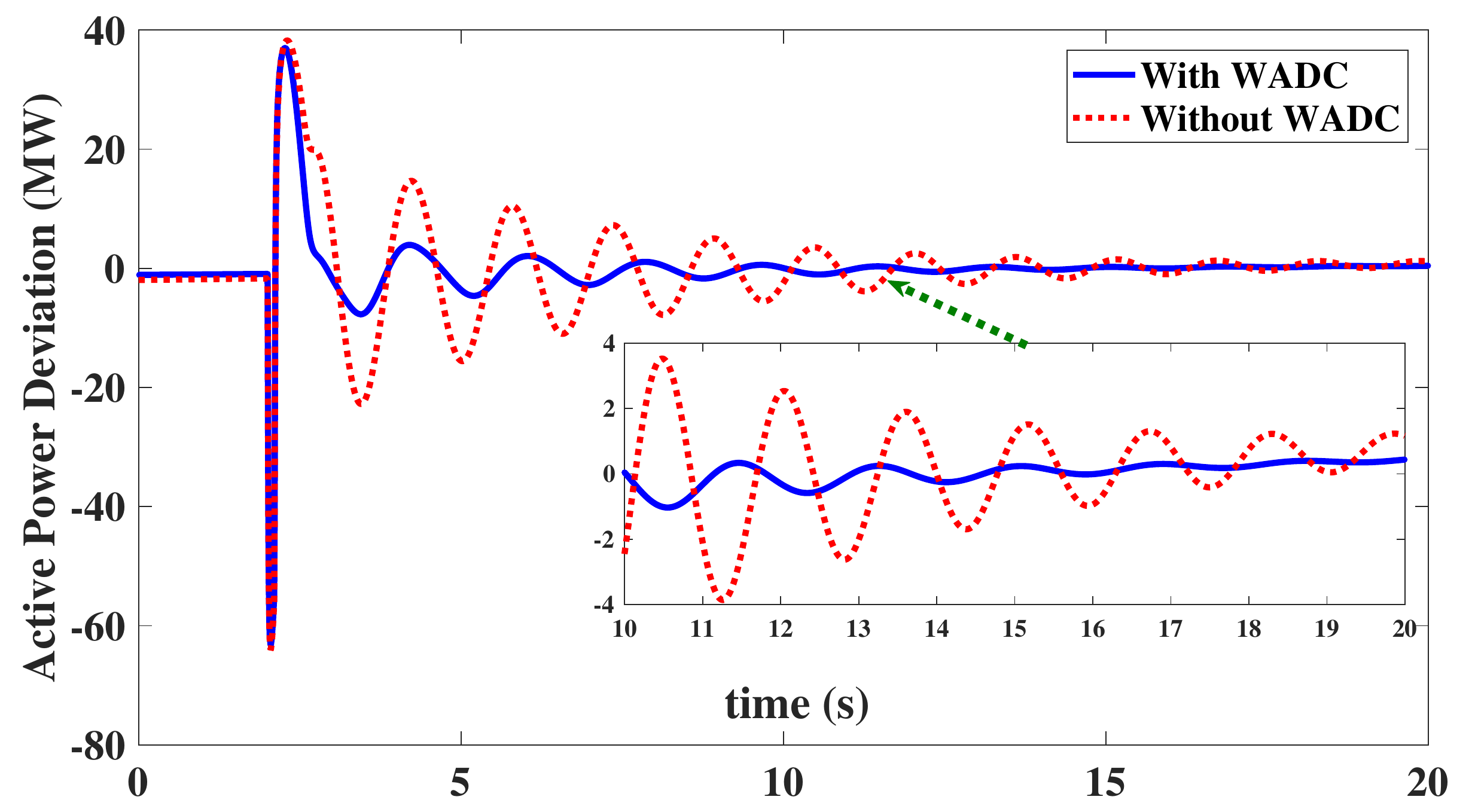}
\caption{Tie-line active power flow deviation ($\Delta P_1$)}
\label{figsx3}
\end{figure}

\begin{figure}[h]
\centering
\includegraphics[width=3.5in]{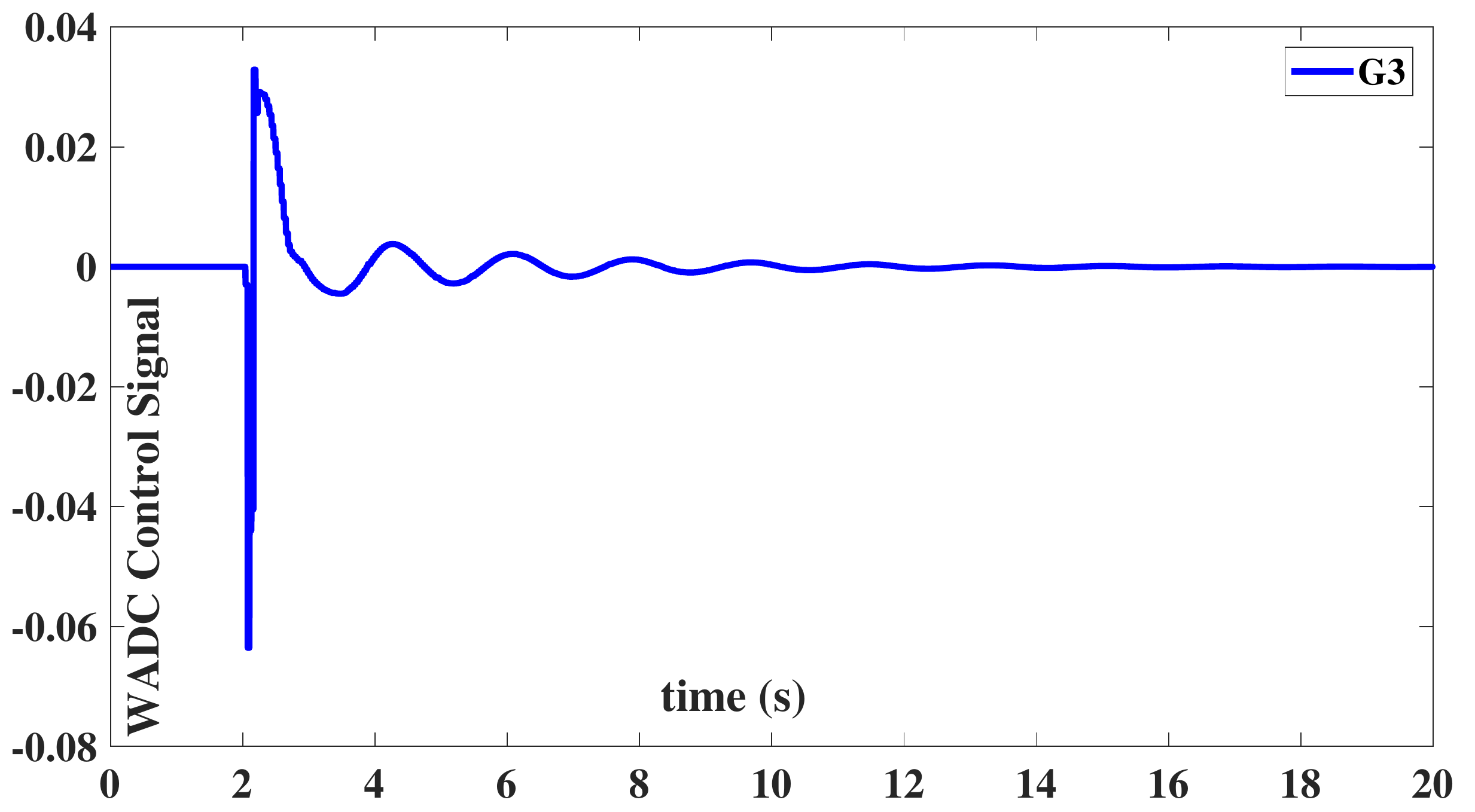}
\caption{WADC output signal}
\label{figs4}
\end{figure}

The wide area control loop for two-area system is as shown in Fig. \ref{figs1}. Fig. \ref{figs2} shows the speed deviation of generator 2 with and without WADC. Fig. \ref{figs3} shows the speed deviation of generator 4 with and without WADC. Fig. \ref{figsx3} shows the tie-line active power deviation $\Delta P_1$. Fig. \ref{figs4} shows the WADC output sent to generator 3. From the above figures it can be seen that the with the proposed approach the oscillations are damped effectively. Table \ref{tab6} shows the comparison of performance metrics with and without WADC. The speed deviation value for generator-2 without WADC at 11s after disturbance is 0.01773 whereas with WADC the value is 0.0038. This indicates that there is 78.68\% reduction in oscillation magnitude. In the same way for generator-4  there is 72.60\% reduction in oscillation magnitude. Further area-under the curve for the absolute value of oscillations is analyzed as shown in Table \ref{tab7} and found that for generator-2 the area under the curve is 66\% less with WADC and for generator-4 it is 56\% less. The area under the curve for tie-line power flow deviation is 51\% less with WADC.

\begin{table}[h]
\processtable{Performance metrics for two-area system \label{tab6}}
{\begin{tabular*}{20pc}{@{\extracolsep{\fill}}lll@{}}\toprule
Variable & Without WADC & With WADC\\
\midrule
Peak $\Delta\omega_2$ at 15s (Fig. \ref{figs2}) & 0.01773 & 0.0038 \\
\hline
Peak $\Delta\omega_4$ at 14.25s (Fig. \ref{figsx3}) & 0.0413 & 0.0113 \\
\botrule
\end{tabular*}}{}
\end{table}

\begin{table}[h]
\processtable{Area under the curve for two-area system \label{tab7}}
{\begin{tabular*}{20pc}{@{\extracolsep{\fill}}lll@{}}\toprule
Variable & Without WADC & With WADC\\
\midrule
$\Delta\omega_2$ (Fig. \ref{figs2}) & 0.4817 & 0.1638 \\
\hline
$\Delta\omega_4$ (Fig. \ref{figs3}) & 0.6959 & 0.3012 \\
\hline
$\Delta P_1$ (Fig. \ref{figsx3}) & 11.739 & 5.75 \\
\botrule
\end{tabular*}}{}
\end{table}

\subsubsection{Effect of time-delay on two-area system control}
\begin{figure}[h]
\centering
\includegraphics[width=3.5in]{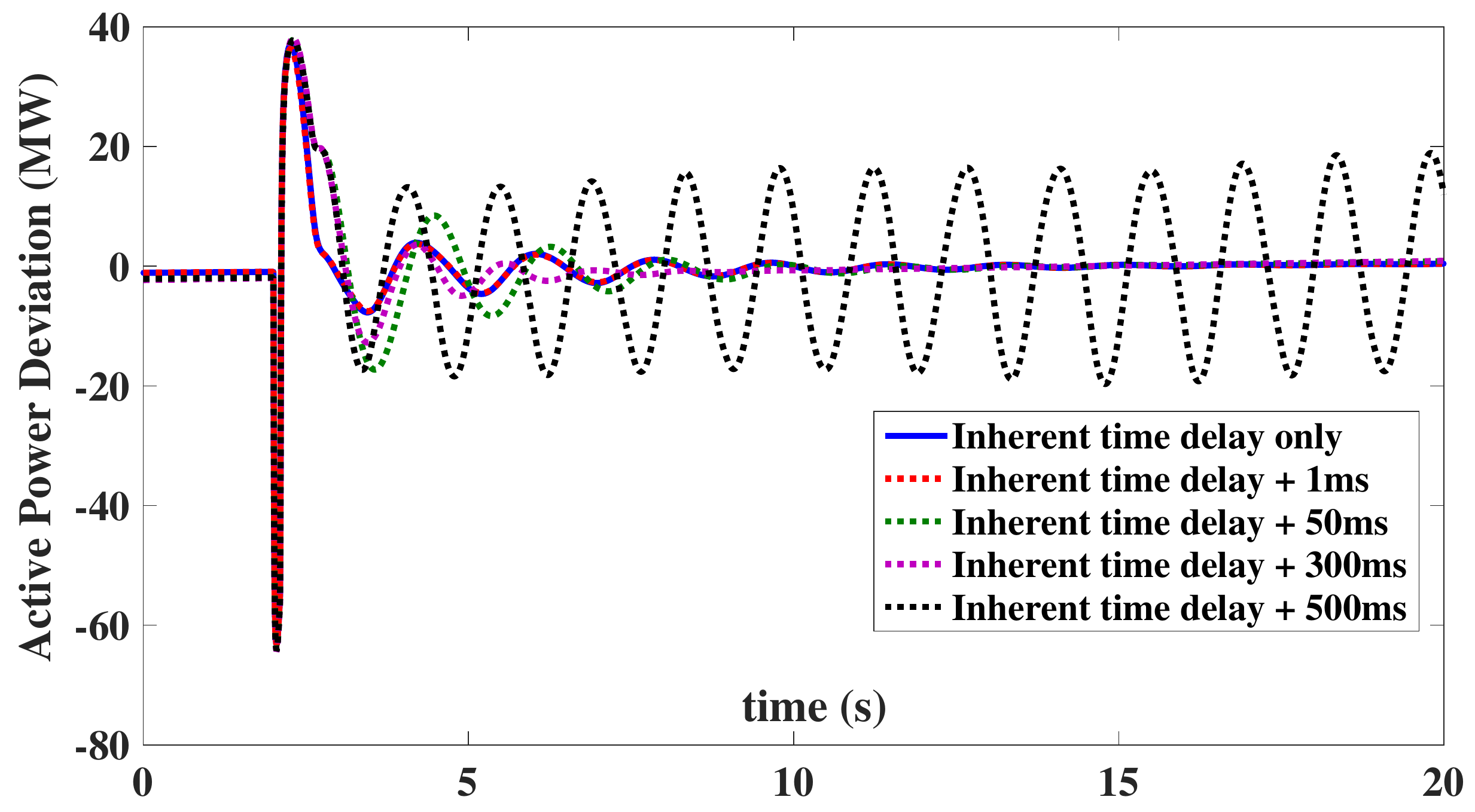}
\caption{Tie-line active power flow deviation $(\Delta P_1)$}
\label{figsx4}
\end{figure}

To study the effect of time-delay on the control, an intentional time-delay is created in addition to the inherent time delay. The inherent time delay for the two-area system case is 0.15s. The intentional time delay is created in the control loop at the time delay block as shown in Fig. \ref{fig1aa}. Initially, a time delay of $1ms$ is created which is further increased up to $500ms$. It can be observed from Fig. \ref{figsx4} that for an extra time delay of $300ms$ the controller performance started to deteriorate and as the time delay is increased further the system is driven into unstable mode. For quantitative analysis, relative error is calculated using \eqref{eqnx44}.
\begin{equation}
\label{eqnx44}
relative\ error = \frac{\left\|\Delta P_{ref}-\Delta P_{delay}\right\|_2}{\left\|\Delta P_{ref}\right\|_2}
\end{equation}
where $\Delta P_{ref}$ represents the response with inherent time delay and $\Delta P_{delay}$  represents the response with intentional time delay. Table \ref{tabxx8} shows the relative error comparison for different time delays. It can be seen that the error increases as the time delay increases.
\begin{table}[h]
\processtable{Relative error for various time delays (two-area system) \label{tabxx8}}
{\begin{tabular*}{20pc}{@{\extracolsep{\fill}}ll@{}}\toprule
Time Delay & Relative Error\\
\midrule
1ms & 0.029319 \\
\hline
50ms & 0.42091 \\
\hline
300ms & 0.52231 \\
\hline
500ms & 1.7834 \\
\botrule
\end{tabular*}}{}
\end{table}

\subsection{Simulation results of 39 bus system}
For this system two cases are analyzed for different disturbances. 

\subsubsection{Case:1}
\begin{figure}[h]
\centering
\includegraphics[width=3.5in]{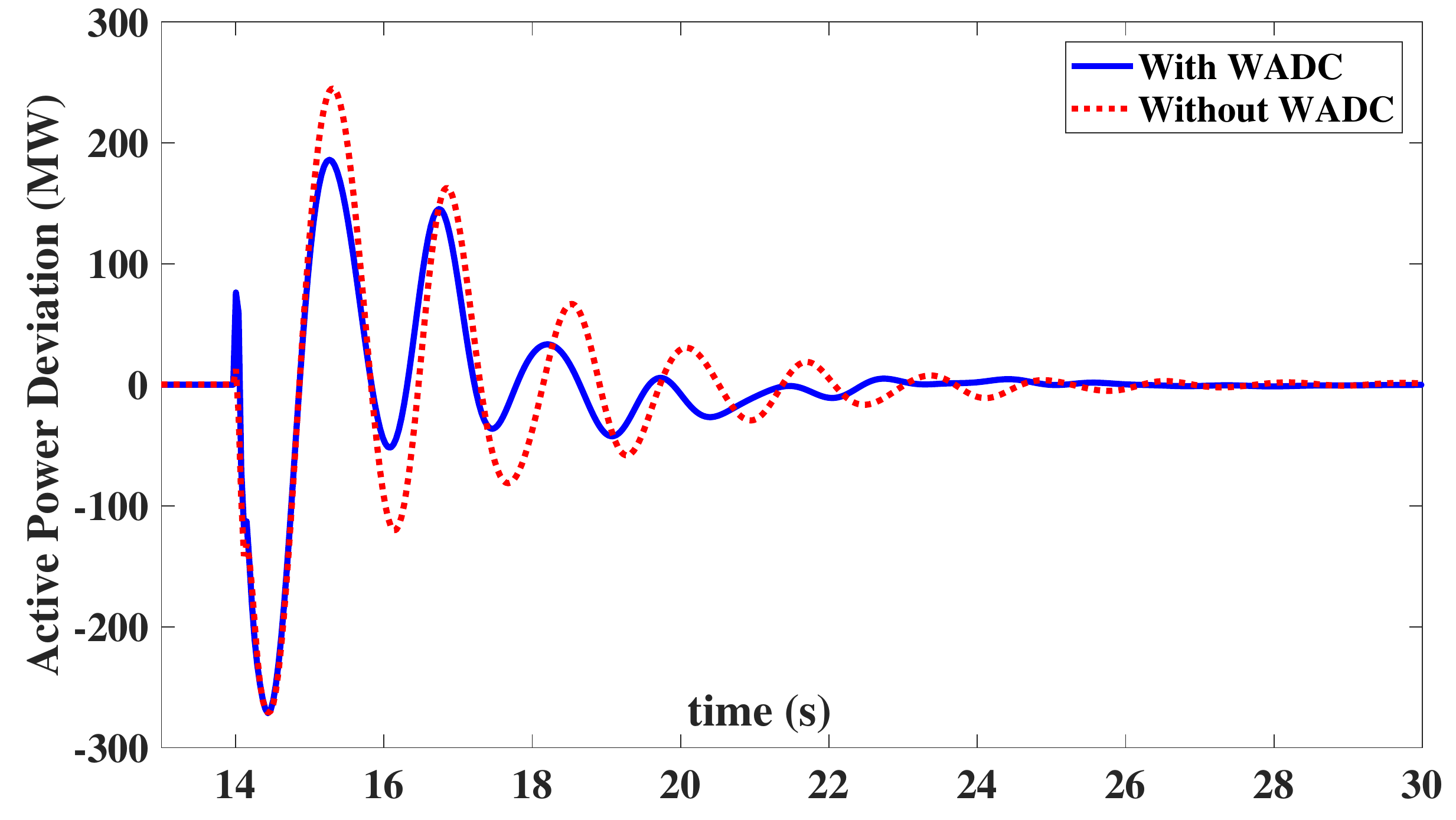}
\caption{Tie-line active power flow deviation ($\Delta P_2\text{ (Bus-39 to Bus-9)}$)}
\label{figz5}
\end{figure}

\begin{figure}[h]
\centering
\includegraphics[width=3.5in]{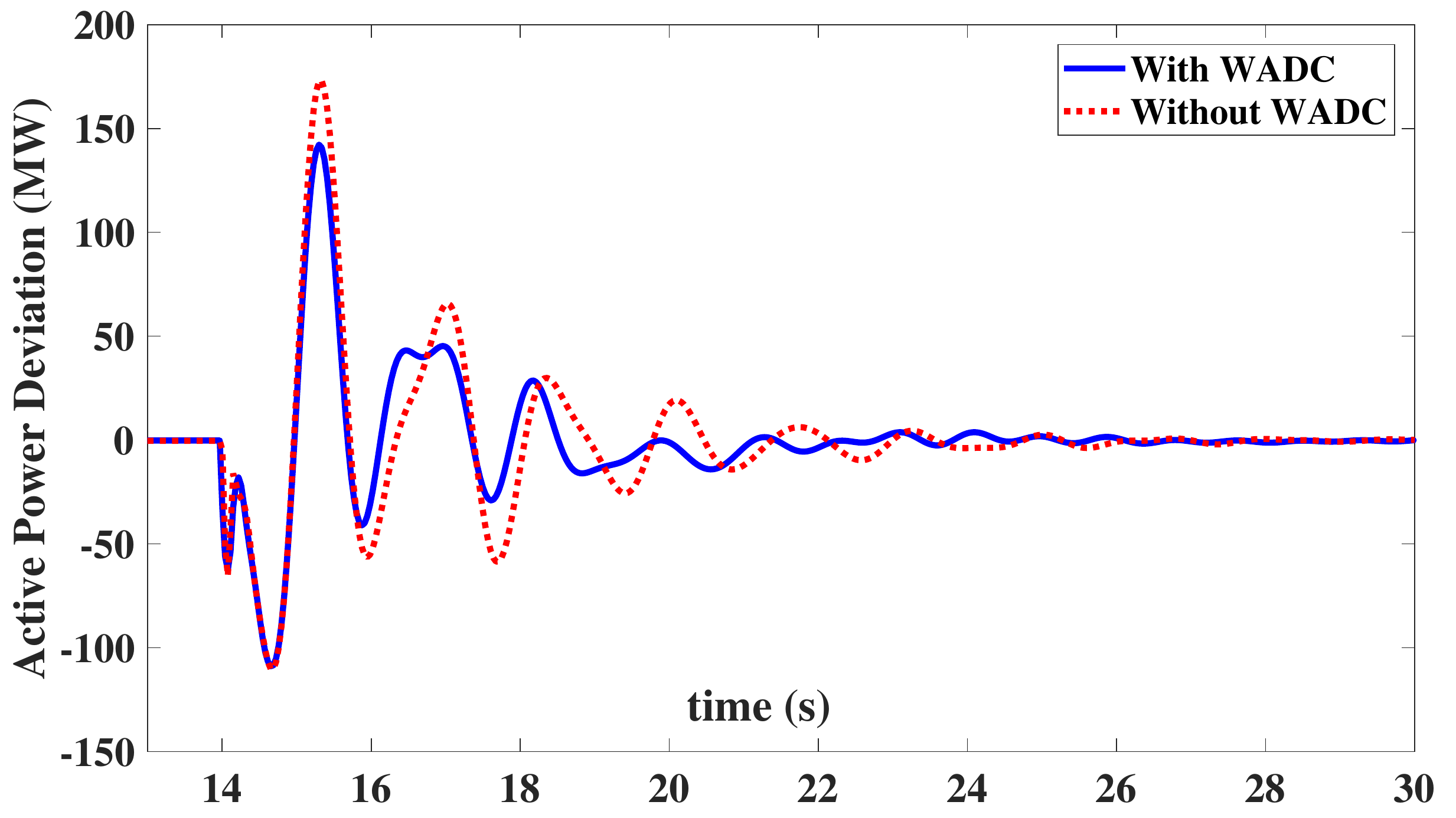}
\caption{Tie-line active power flow deviation ($\Delta P_4\text{ (Bus-14 to Bus-15)}$)}
\label{figz6}
\end{figure}

\begin{figure}[h]
\centering
\includegraphics[width=3.5in]{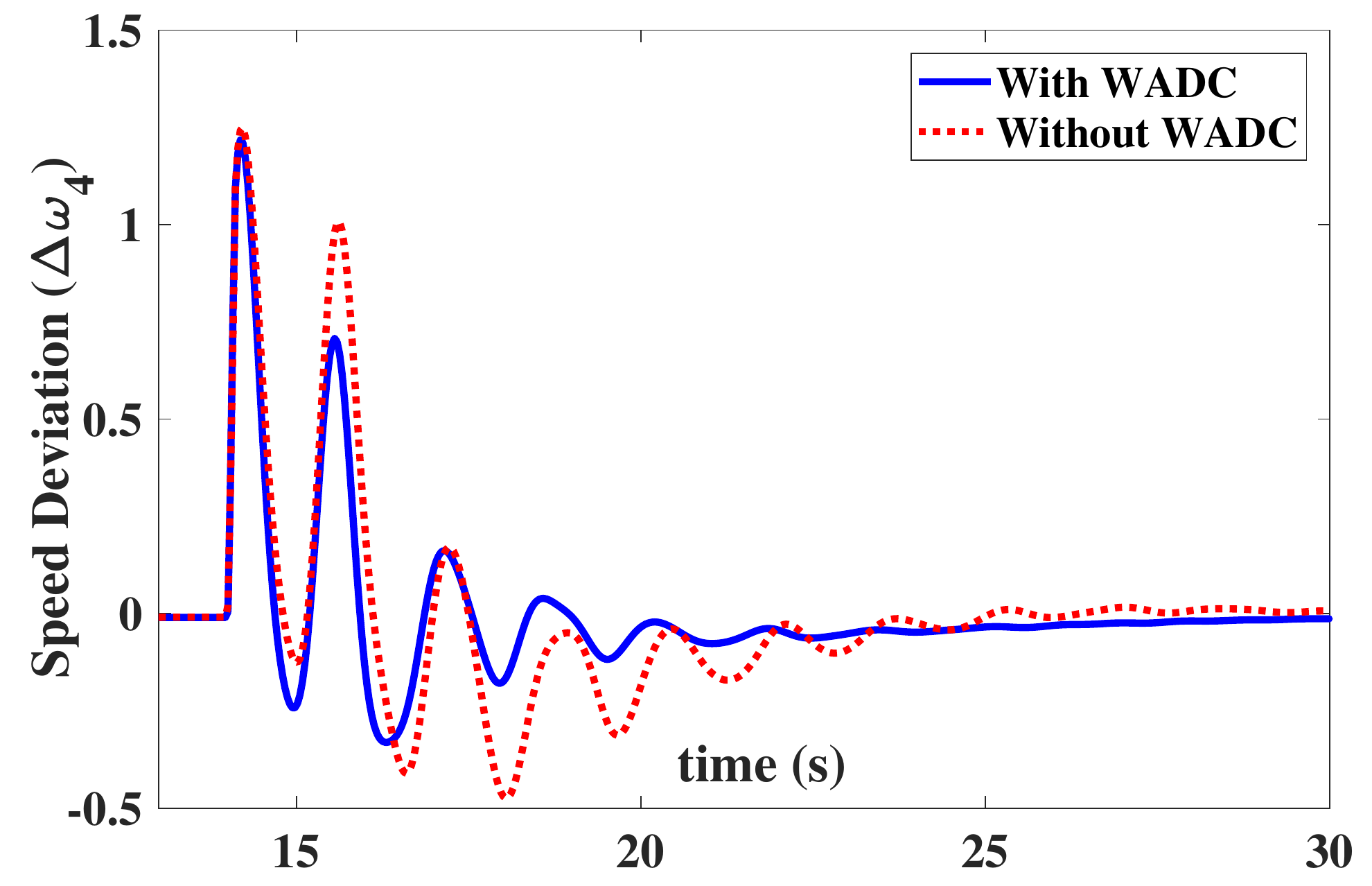}
\caption{Generator-4 speed deviation (rad/s)}
\label{figs6}
\end{figure}

\begin{figure}[h]
\centering
\includegraphics[width=3.5in]{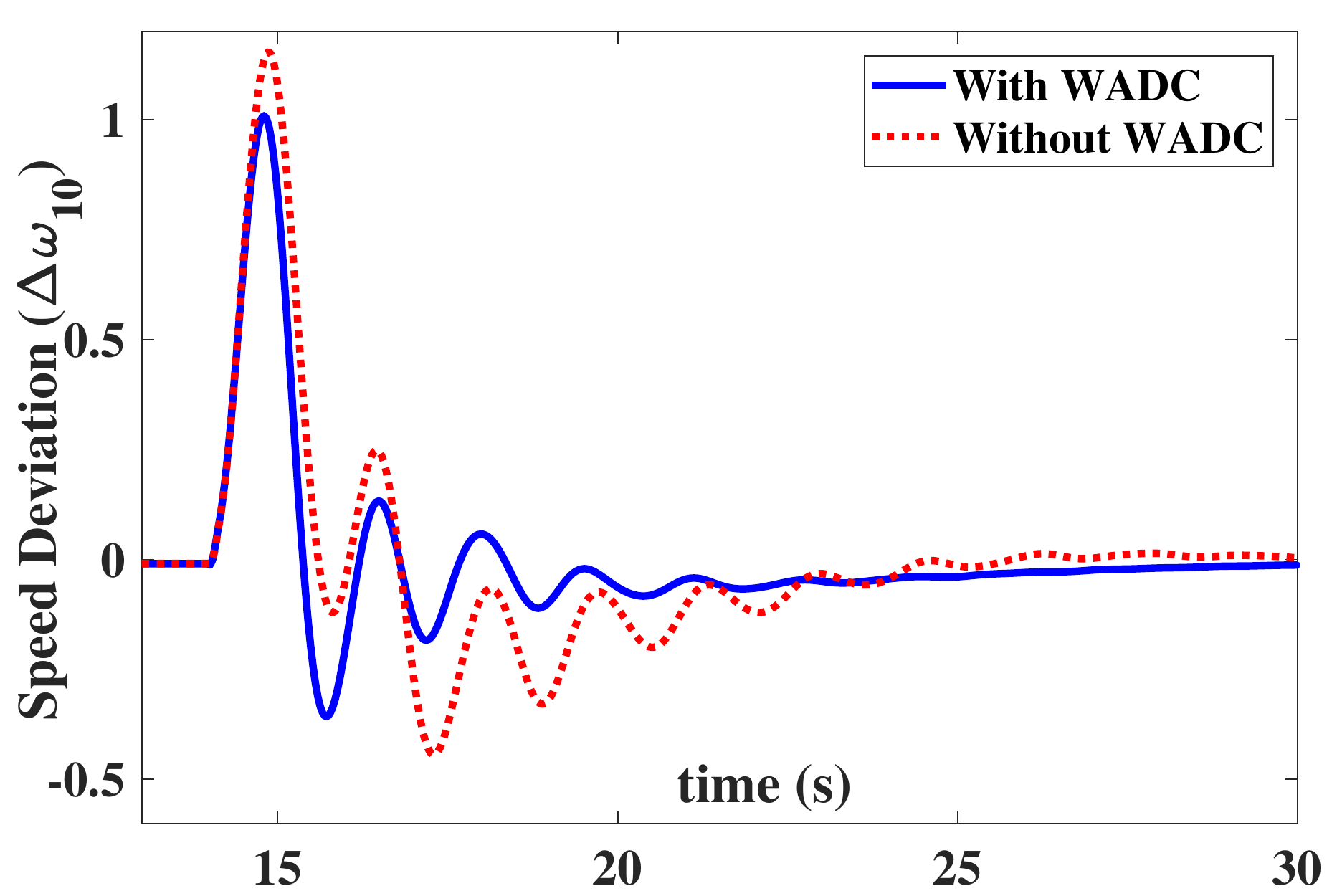}
\caption{Generator-10 speed deviation (rad/s)}
\label{figs8}
\end{figure}

\begin{figure}[h]
\centering
\includegraphics[width=3.5in]{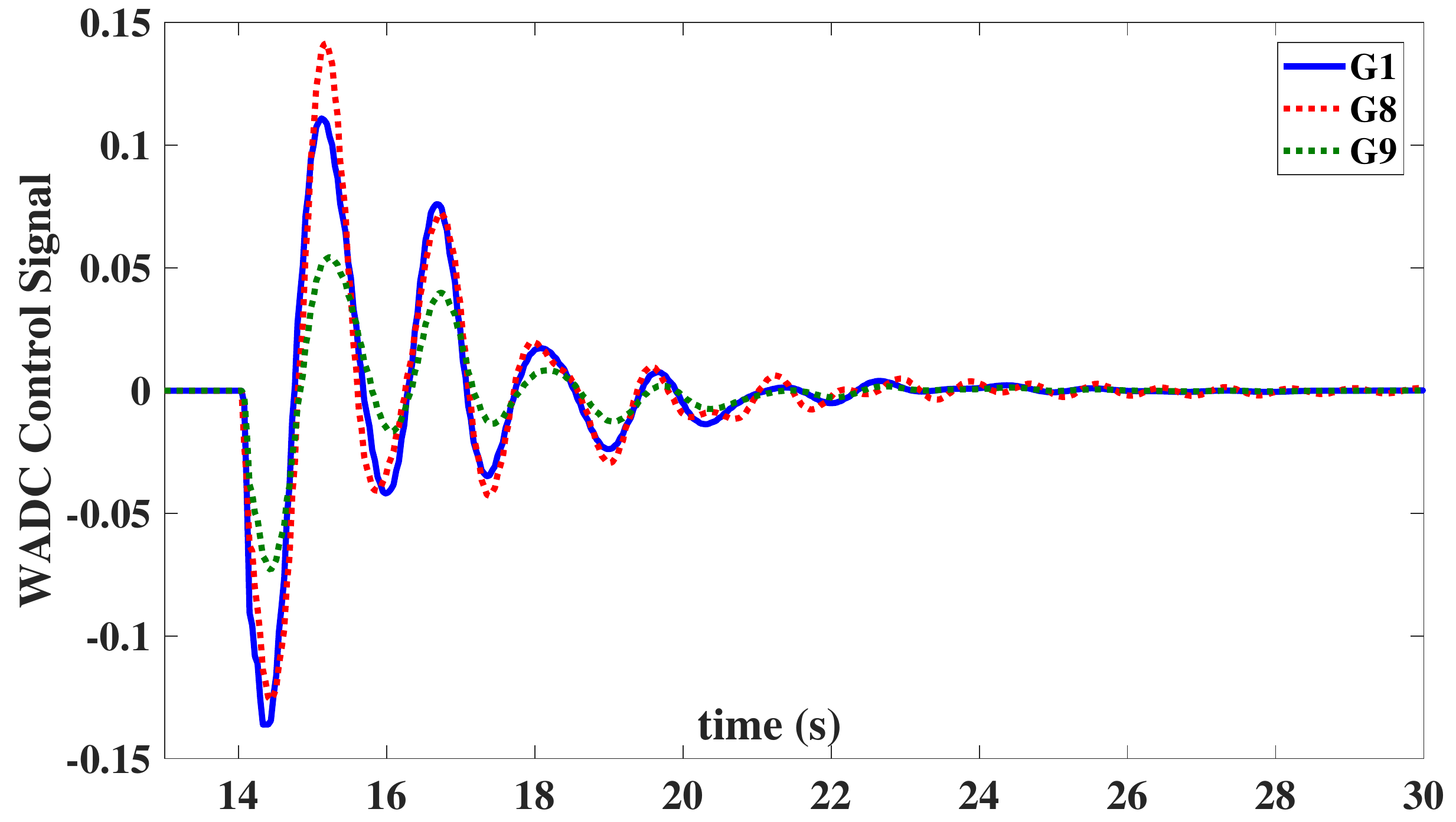}
\caption{WADC output signal}
\label{figs9}
\end{figure}

For the same disturbance scenario as discussed in Section \ref{c1}, Fig. \ref{figz5} and Fig. \ref{figz6} shows the active power deviations $\Delta P_2$, and $\Delta P_4$ respectively. Fig. \ref{figs6}, and Fig. \ref{figs8} shows the speed deviations of generator 4, and 10 with and without WADC respectively. Fig. \ref{figs9} shows the WADC output sent to most controllable generator. From the above figures it can be seen that with the proposed approach oscillations are damped effectively. The area under the curve for absolute value of speed deviations are analyzed as shown in Table \ref{tab8} and it is found that the area under the curve is reduced by 59\%, and 60.56\% with WADC for generators 4, and 10 respectively.

\begin{table}[h]
\processtable{Area under the curve metric (39 bus system case:1) \label{tab8}}
{\begin{tabular*}{20pc}{@{\extracolsep{\fill}}lll@{}}\toprule
Variable & Without WADC & With WADC\\
\midrule
$\Delta\omega_4$ (Fig. \ref{figs6}) & 6.3374 & 2.7498 \\
\hline
$\Delta\omega_{10}$ (Fig. \ref{figs8}) & 6.3081 & 2.4878 \\
\botrule
\end{tabular*}}{}
\end{table}

\begin{figure}[h]
\centering
\includegraphics[width=3.5in]{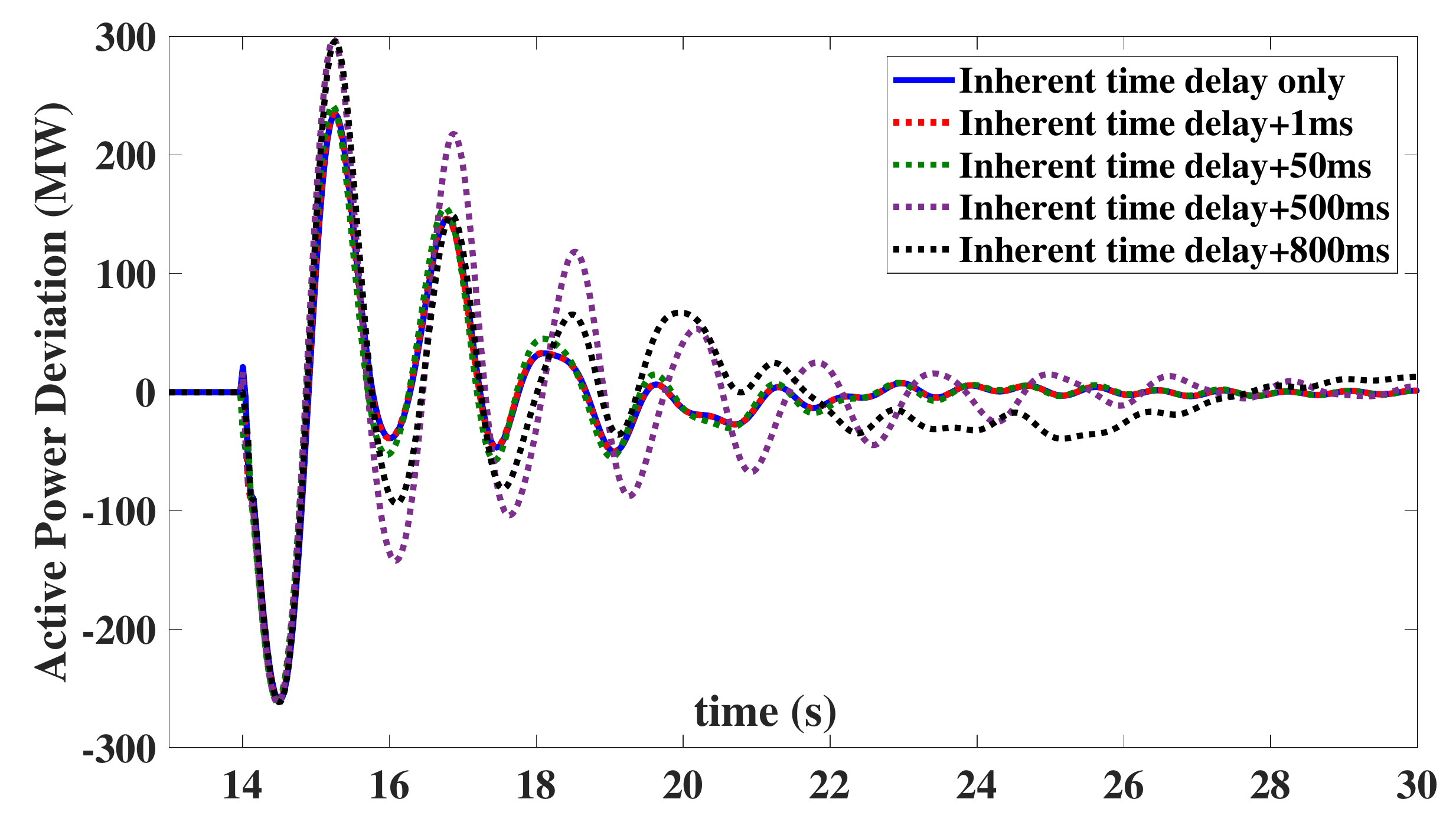}
\caption{Tie-line active power flow deviation $(\Delta P_1)$}
\label{figsx9}
\end{figure}

\subsubsection{Case:2}
\begin{figure}[h]
\centering
\includegraphics[width=3.5in]{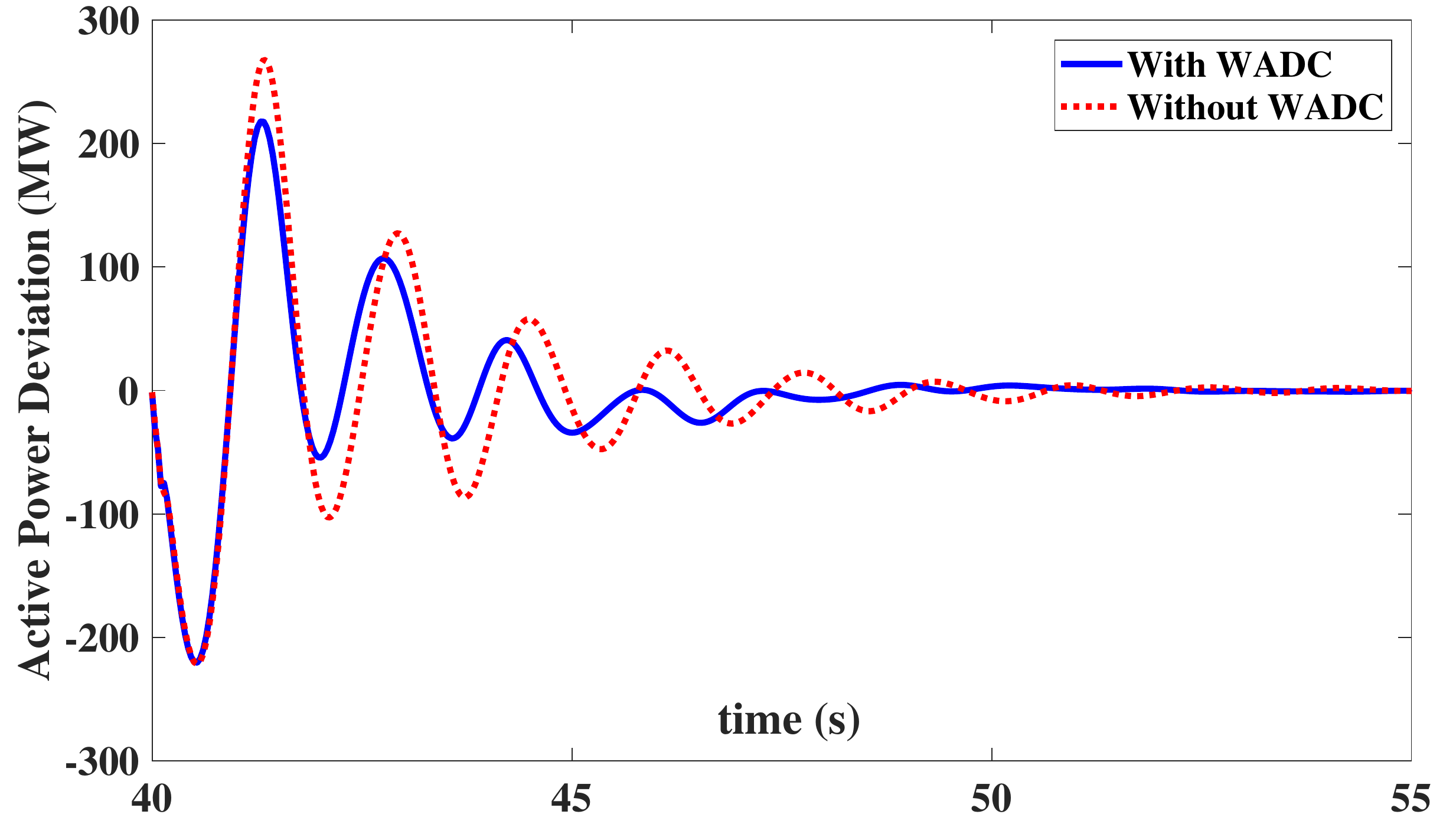}
\caption{Tie-line active power flow deviation ($\Delta P_2\text{ (Bus-39 to Bus-9)}$)}
\label{figz10}
\end{figure}

\begin{figure}[h]
\centering
\includegraphics[width=3.5in]{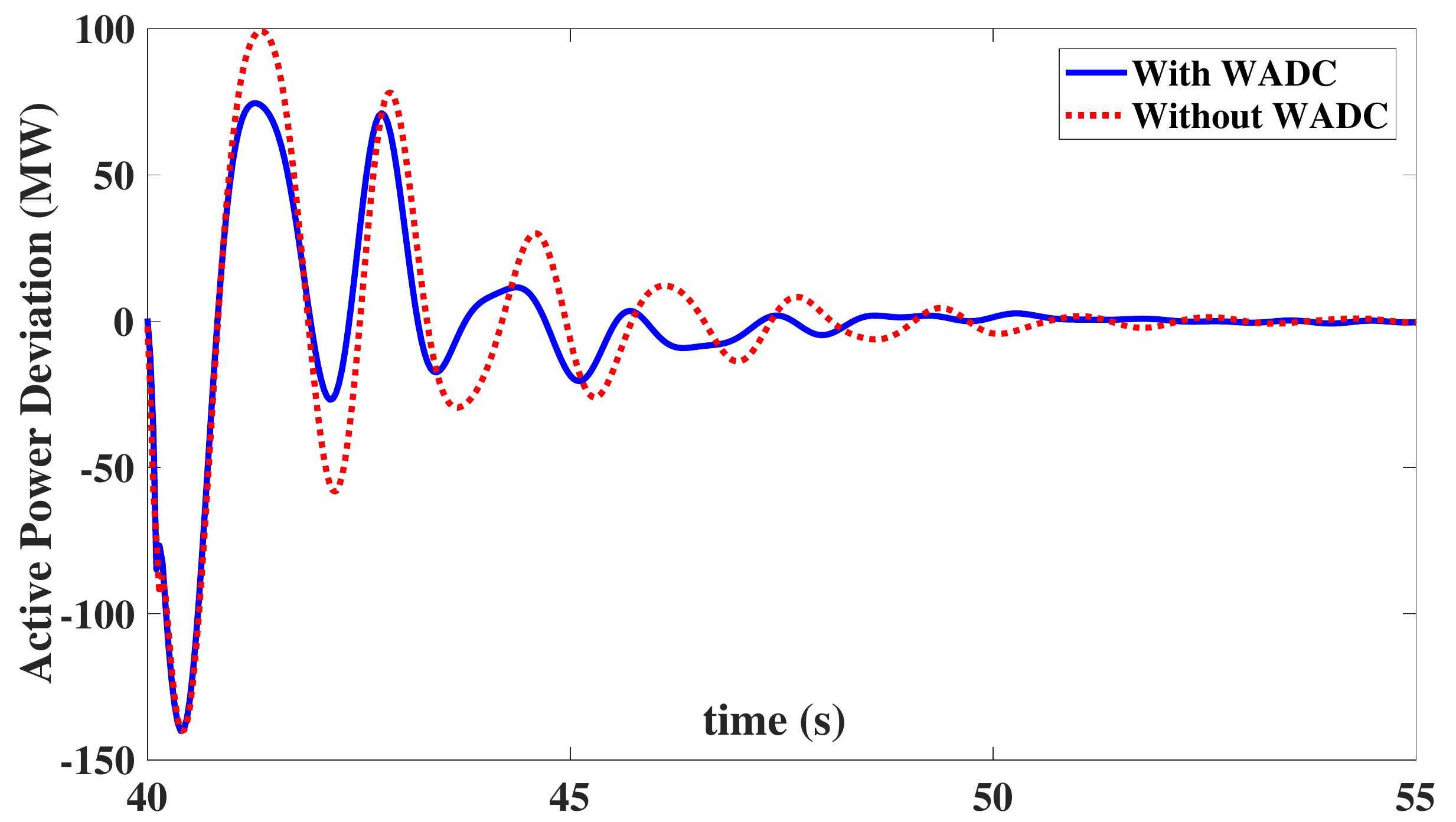}
\caption{Tie-line active power flow deviation ($\Delta P_4\text{ (Bus-14 to Bus-15)}$)}
\label{figz11}
\end{figure}

\begin{figure}[h]
\centering
\includegraphics[width=3.5in]{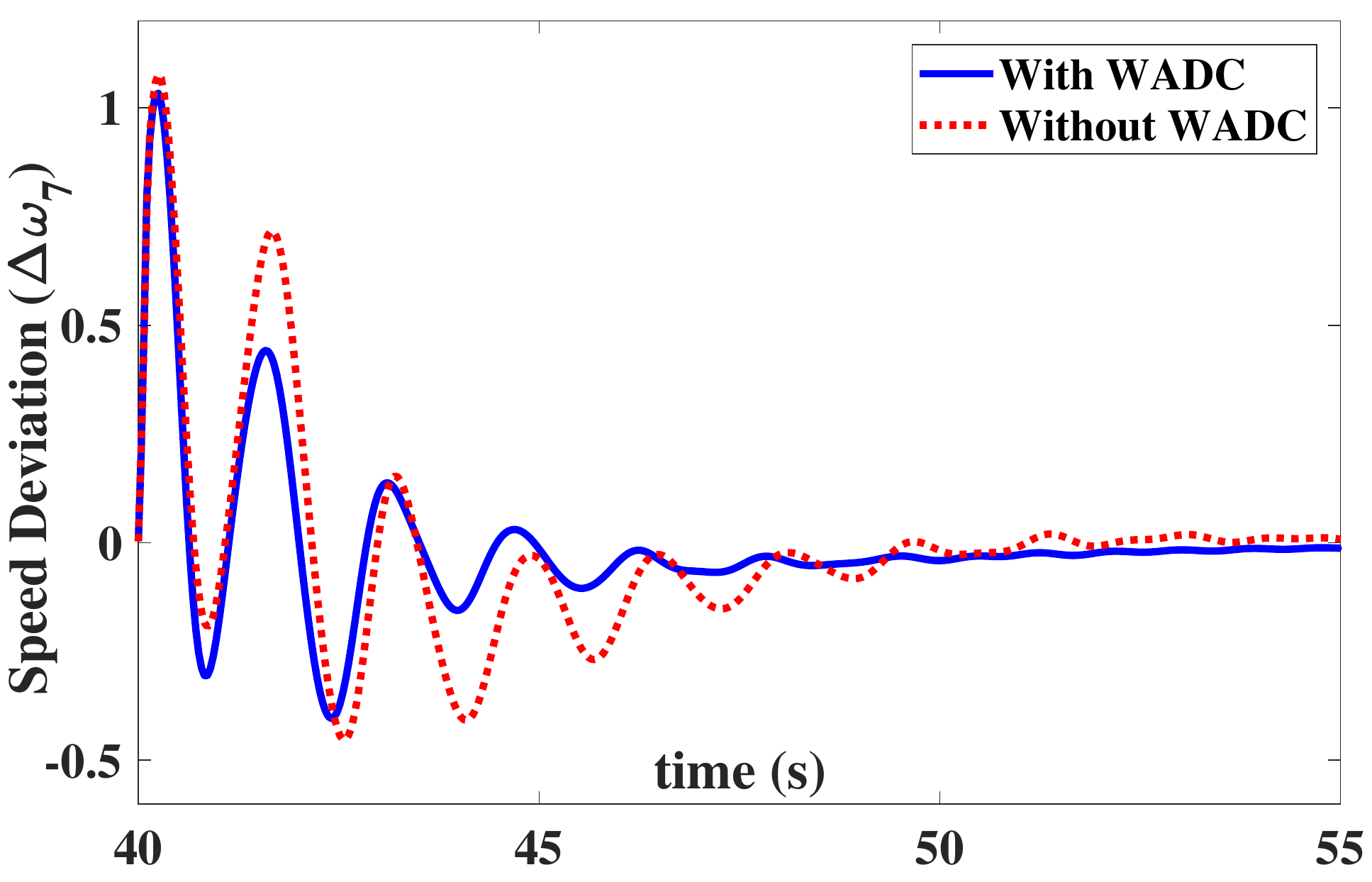}
\caption{Generator-7 speed deviation (rad/s)}
\label{figs12}
\end{figure}

\begin{figure}[h]
\centering
\includegraphics[width=3.5in]{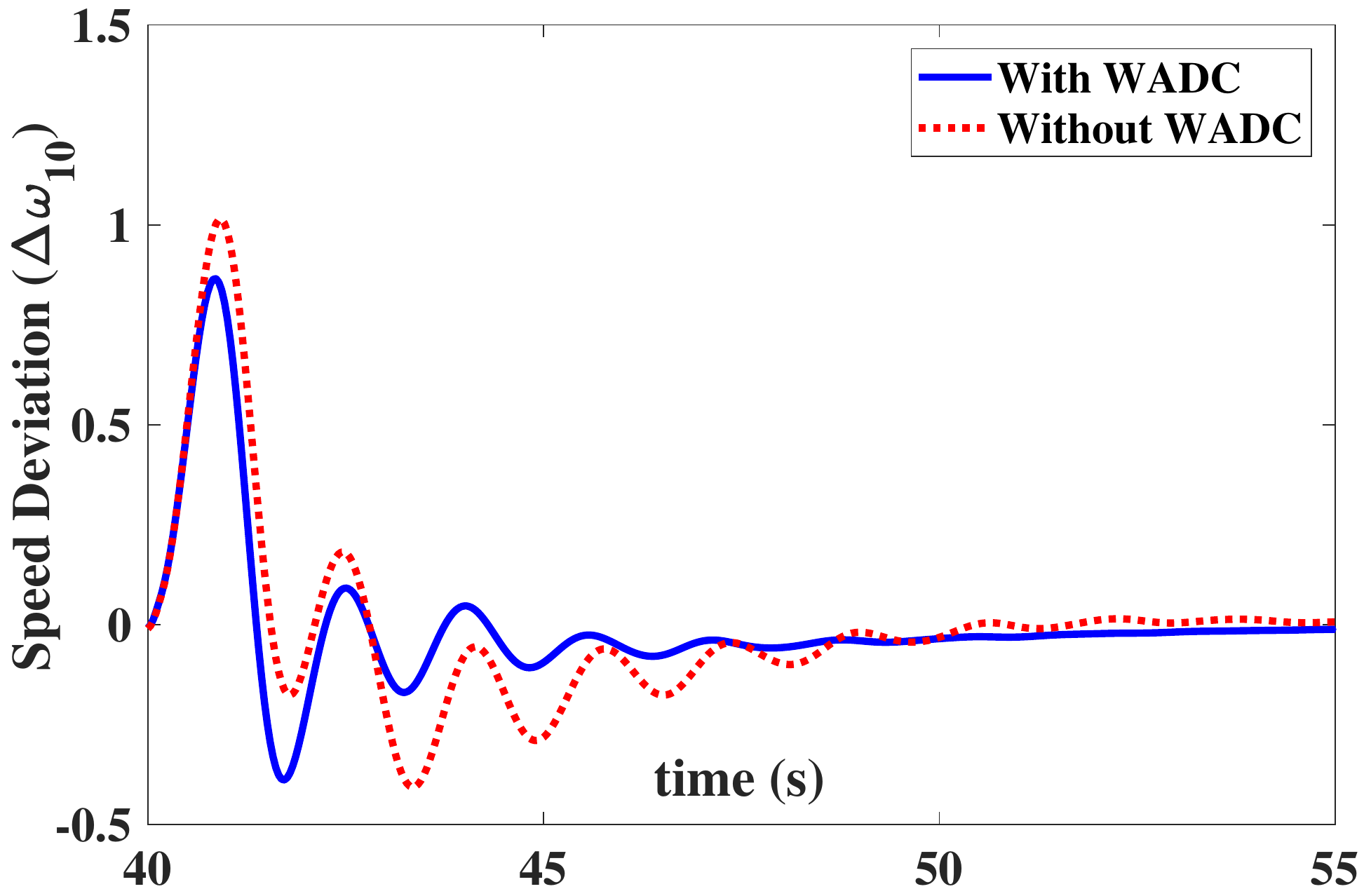}
\caption{Generator-10 speed deviation (rad/s)}
\label{figs13}
\end{figure}

\begin{figure}[h]
\centering
\includegraphics[width=3.5in]{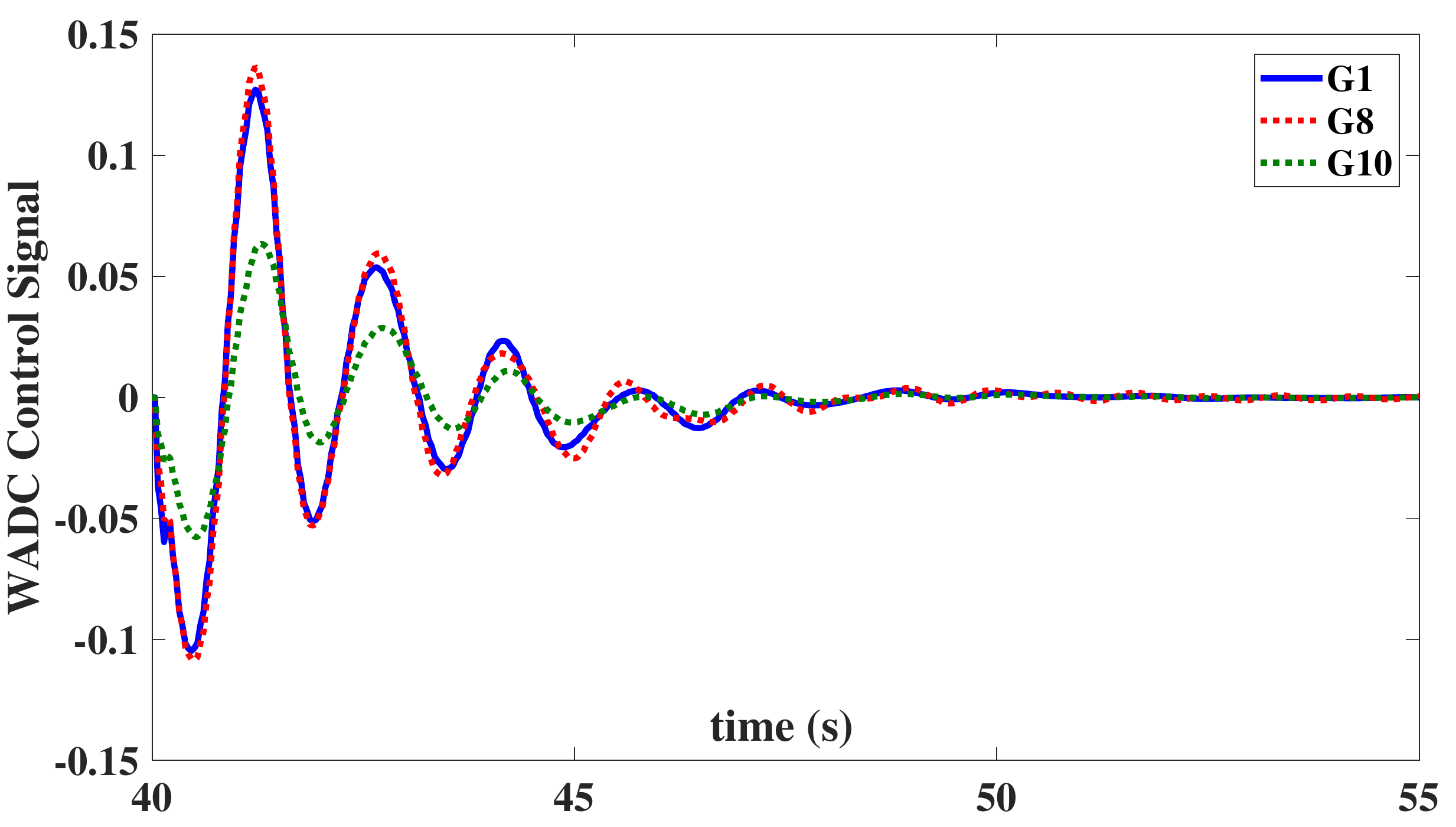}
\caption{WADC output signal}
\label{figs14}
\end{figure}

For the same disturbance scenario as discussed in Section \ref{c2}. Fig. \ref{figz10} and Fig. \ref{figz11} shows the active power deviations $\Delta P_2$, and $\Delta P_4$ respectively. Fig. \ref{figs12}, and Fig. \ref{figs13} shows the speed deviations of generator 7, and 10 with and without WADC respectively. Fig. \ref{figs14} shows the WADC output sent to most controllable generator.  From the above mentioned figures it can be seen that the with the proposed approach the oscillations are damped effectively. The area under the curve for absolute value of speed deviations are analyzed as shown in Table \ref{tab9} and it is found that the area under the curve is reduced by 54\%, and 56\% with WADC for generators 7, and 10 respectively.

\begin{table}[h]
\processtable{Area under the curve metric (39 bus system case:2) \label{tab9}}
{\begin{tabular*}{20pc}{@{\extracolsep{\fill}}lll@{}}\toprule
Variable & Without WADC & With WADC\\
\midrule
$\Delta\omega_7$ (Fig. \ref{figs12}) & 3.746 & 1.7233 \\
\hline
$\Delta\omega_{10}$ (Fig. \ref{figs13}) & 3.5383 & 1.5359 \\
\botrule
\end{tabular*}}{}
\end{table}

\begin{figure}[h]
\centering
\includegraphics[width=3.5in]{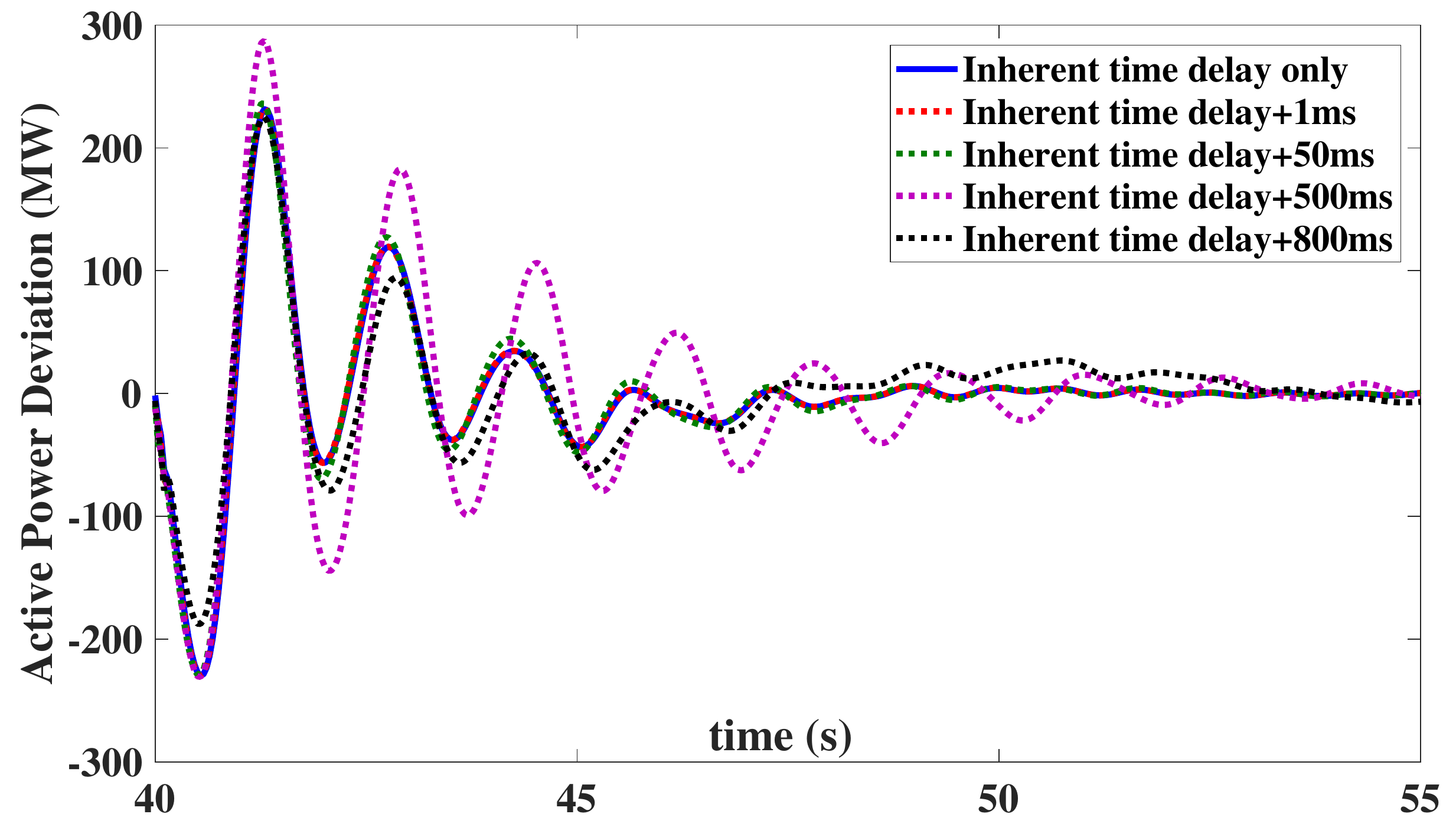}
\caption{Tie-line active power flow deviation $(\Delta P_1)$}
\label{figsx14}
\end{figure}

\subsubsection{Effect of time-delay on 39 bus system control}
The inherent time delay for the 39 bus system case is 0.23s. The intentional time delay is created in the control loop at the time delay block as shown in Fig. \ref{fig1aa}. Initially, a time delay of $1ms$ is created and increased up to $800ms$. It can be observed from Fig. \ref{figsx9} and Fig. \ref{figsx14} that for an extra time delay of $500ms$ the controller performance started to deteriorate and as the time delay is increased further the system oscillations also increased. Table \ref{tabxx9} shows the relative error comparison for 39 bus system for various time delays calculated using \eqref{eqnx44}. In this case also the relative error increases as the time delay increases.

\begin{table}[h]
\processtable{Relative error for various time delays (39 bus system) \label{tabxx9}}
{\begin{tabular*}{20pc}{@{\extracolsep{\fill}}lll@{}}\toprule
 & Case:1 & Case:2 \\
\midrule
Time Delay & Relative Error & Relative Error\\
\hline
1ms & 0.0526 & 0.0376\\
\hline
50ms & 0.1469 & 0.1311\\
\hline
300ms & 0.4979 & 0.3216\\
\hline
500ms & 0.6429 & 0.6418\\
\botrule
\end{tabular*}}{}
\end{table}

\section{Conclusion}\label{sec10}
The proposed method for measurement based wide area damping of inter-area oscillations based on MIMO identification overcome the drawbacks of earlier linearization based methods reported in literature. In this approach, initially MIMO system transfer functions are identified, then the inter-area modes are estimated using the MIMO transfer functions. The optimal control loop required for WADC is estimated through residues corresponding to the inter-area mode of interest. Finally, the WADC is design based on a combination of DLQR and Kalman filtering algorithms. The effectiveness of the proposed algorithm is verified by implementing on two-area and IEEE 39 bus power system models on RTDS/RSCAD and MATLAB based real-time co-simulation platform. In the future, framework to mitigate time-delay effects on the control action will be addressed.

\bibliographystyle{iet}
\bibliography{Main.bbl}

\vfill\pagebreak

\end{document}